%% file: body.tex
\documentclass[twocolumn]{pasj00}
\usepackage{lscape}
\usepackage[normalem]{ulem}

\begin{document}
\SetRunningHead{K. Hiroi et al.}
{The First MAXI/GSC Catalog in the High Galactic-Latitude Sky}
\Received{2011/06/12}
\Accepted{2011/08/13}
\Published{}

\title{The First MAXI/GSC Catalog in the High Galactic-Latitude Sky}

\author{
Kazuo \textsc{Hiroi},\altaffilmark{1}
Yoshihiro \textsc{Ueda},\altaffilmark{1}
Naoki \textsc{Isobe},\altaffilmark{1,2}
Masaaki \textsc{Hayashida},\altaffilmark{1}
Satoshi \textsc{Eguchi},\altaffilmark{1,3}
Mutsumi \textsc{Sugizaki},\altaffilmark{4}
Nobuyuki \textsc{Kawai},\altaffilmark{5}
Hiroshi \textsc{Tsunemi},\altaffilmark{6}
Masaru \textsc{Matsuoka},\altaffilmark{4,7}
Tatehiro \textsc{Mihara},\altaffilmark{4}
Kazutaka \textsc{Yamaoka},\altaffilmark{8}
Masaki \textsc{Ishikawa},\altaffilmark{9}
Masashi \textsc{Kimura},\altaffilmark{6}
Hiroki \textsc{Kitayama},\altaffilmark{6}
Mitsuhiro \textsc{Kohama},\altaffilmark{7}
Takanori \textsc{Matsumura},\altaffilmark{10}
Mikio \textsc{Morii},\altaffilmark{5}
Yujin E. \textsc{Nakagawa},\altaffilmark{11}
Satoshi \textsc{Nakahira},\altaffilmark{4}
Motoki \textsc{Nakajima},\altaffilmark{12}
Hitoshi \textsc{Negoro},\altaffilmark{13}
Motoko \textsc{Serino},\altaffilmark{4}
Megumi \textsc{Shidatsu},\altaffilmark{1}
Tetsuya \textsc{Sootome},\altaffilmark{4}
Kousuke \textsc{Sugimori},\altaffilmark{5}
Fumitoshi \textsc{Suwa},\altaffilmark{13}
Takahiro \textsc{Toizumi},\altaffilmark{5}
Hiroshi \textsc{Tomida},\altaffilmark{7}
Yohko \textsc{Tsuboi},\altaffilmark{10}
Shiro \textsc{Ueno},\altaffilmark{7}
Ryuichi \textsc{Usui},\altaffilmark{5}
Takayuki \textsc{Yamamoto},\altaffilmark{4}
Kyohei \textsc{Yamazaki},\altaffilmark{10}
Atsumasa \textsc{Yoshida},\altaffilmark{8}
and the MAXI team} 
\altaffiltext{1}{Department of Astronomy, Kyoto University, Oiwake-cho, Sakyo-ku, Kyoto 606-8502}
\email{hiroi@kusastro.kyoto-u.ac.jp}
\altaffiltext{2}{Institute of Space and Astronautical Science (ISAS), Japan Aerospace Exploration Agency (JAXA) ,3-1-1 Yoshino-dai, Chuo-ku, Sagamihara, Kanagawa 252-5210}
\altaffiltext{3}{National Astronomical Observatory of Japan, 2-21-1, Osawa, Mitaka City, Tokyo 181-8588}
\altaffiltext{4}{MAXI team, Institute of Physical and Chemical Research (RIKEN), 2-1 Hirosawa, Wako, Saitama 351-0198}
\altaffiltext{5}{Department of Physics, Tokyo Institute of Technology, 2-12-1 Ookayama, Meguro-ku, Tokyo 152-8551}
\altaffiltext{6}{Department of Earth and Space Science, Osaka University, 1-1 Machikaneyama, Toyonaka, Osaka 560-0043}
\altaffiltext{7}{ISS Science Project Office, Institute of Space and Astronautical Science (ISAS), Japan Aerospace Exploration Agency (JAXA), 2-1-1 Sengen, Tsukuba, Ibaraki 305-8505}
\altaffiltext{8}{Department of Physics and Mathematics, Aoyama Gakuin University,\\ 5-10-1 Fuchinobe, Chuo-ku, Sagamihara, Kanagawa 252-5258}
\altaffiltext{9}{School of Physical Science, Space and Astronautical Science, The graduate University for Advanced Studies (Sokendai), Yoshinodai 3-1-1, Chuo-ku, Sagamihara, Kanagawa 252-5210}
\altaffiltext{10}{Department of Physics, Chuo University, 1-13-27 Kasuga, Bunkyo-ku, Tokyo 112-8551}
\altaffiltext{11}{Research Institute for Science and Engineering, Waseda University, 17 Kikui-cho, Shinjuku-ku, Tokyo 162-0044}
\altaffiltext{12}{School of Dentistry at Matsudo, Nihon University, 2-870-1 Sakaecho-nishi, Matsudo, Chiba 101-8308}
\altaffiltext{13}{Department of Physics, Nihon University, 1-8-14 Kanda-Surugadai, Chiyoda-ku, Tokyo 101-8308}



\maketitle

\begin{abstract}

We present the first unbiased source catalog of the Monitor of All-sky
X-ray Image (MAXI) mission at high Galactic latitudes ($|b| >
10^{\circ}$), produced from the first 7-month data (2009 September 1
to 2010 March 31) of the Gas Slit Camera in the 4--10 keV band. We
develop an analysis procedure to detect faint sources from the
MAXI data, utilizing a maximum likelihood image fitting method, where
the image response, background, and detailed observational conditions
are taken into account. The catalog consists of 143 X-ray sources
above 7 sigma significance level with a limiting sensitivity of
$\sim1.5\times10^{-11}$ ergs cm$^{-2}$ s$^{-1}$ (1.2 mCrab) in the
4--10 keV band. Among them, we identify 38 Galactic/LMC/SMC objects,
48 galaxy clusters, 39 Seyfert galaxies, 12 blazars, and 1 galaxy. Other 4
sources are confused with multiple objects, and one remains
unidentified. The log $N$ - log $S$ relation of extragalactic objects is
in a good agreement with the HEAO-1 A-2 result, although the list of
the brightest AGNs in the entire sky has significantly changed since 
that in 30 years ago.

\end{abstract}

\section{INTRODUCTION}
\label{sec:INTRODUCTION}

All-sky X-ray surveys are powerful tools to investigate the whole
populations of active and hot phenomena in the universe at the
brightest flux end. The strong X-ray emitters include Galactic objects
such as active stars, SNRs, pulsars, CVs, low mass and high mass X-ray
binaries (with a neutron star or a black hole as the primary), and
extragalactic objects, mainly active galactic nuclei (AGNs; Seyfert
galaxies and blazars) and clusters of galaxies. The source catalog
consisting of a statistically well-defined sample detected from an
unbiased survey is a primary product on which many subsequent
studies are based. For extragalactic populations, in particular, these
results define the ``local'' sample in the present universe, the end
point of their cosmological evolution. Thus, to establish the
statistical properties of bright X-ray sources 
using the best quality data over the entire sky has always been 
a key issue in high energy astrophysics.

Past all-sky X-ray surveys indeed brought valuable information on the
X-ray source populations. In the soft X-ray band, the ROSAT mission
conducted an all-sky survey in the 0.1--2.4 keV band, producing the
ROSAT All-Sky Survey (RASS) Bright Source Catalog (BSC; \cite{1999A&A...349..389V}) 
and Faint Source Catalog (FSC; \cite{{2000IAUC.7432....3V}}), which 
contain 18,811 and 105,924 sources,
respectively. Because of its large sample size, only a part of the
RASS sources has been optically identified (e.g., \cite{2000AN....321....1S}). 
Hard X-rays
above 2 keV are more effective to detect obscured objects, such as
``type 2'' AGNs, due to its strong penetrating power against
photoelectric absorption. In the late 1970s, HEAO-1 A-2 performed an
all-sky X-ray survey at $|b| > 20^{\circ}$ in the 2--10 keV band down
to a limiting sensitivity of $\sim3.1\times10^{-11}$ ergs cm$^{-2}$
s$^{-1}$, detecting 61 extragalactic sources including 29 AGNs 
(\cite{1982ApJ...253..485P}). {\it Rossi} X-ray Timing
Explorer (RXTE) also carried out an all-sky survey using ``slew'' mode
data of the Proportional Counter Array (PCA) in the 3--8 keV and 8--20
keV bands, and achieved the sensitivity similar to, and an order of
magnitude higher than those of HEAO-1 A-1 and A-4, respectively
(\cite{2004A&A...418..927R}). The RXTE/PCA survey detected 294 sources
including 100 AGNs, although the identification is not complete
(80\%). Recently, the Swift and INTEGRAL satellites have performed
all-sky surveys in the hard X-ray band above 10 keV 
(Swift: \cite{2008ApJ...681..113T}; \cite{2010ApJS..186..378T}; 
\cite{2010A&A...524A..64C}; \cite{Baumgartner2010}, 
INTEGRAL: \cite{2007ApJS..170..175B}; 
Beckmann et al. 2006, 2009; \cite{2007A&A...475..775K}; \cite{2010ApJS..186....1B}).
These catalogs contain heavily obscured AGNs with absorption column
densities larger than 10$^{24}$ cm$^{-2}$. Total 628 AGNs have
been detected in the Palermo Swift/BAT 54-month catalog
(\cite{2010A&A...524A..64C}).

Monitor of All-sky X-ray Image (MAXI; \cite{2009PASJ...61..999M}) is
the first scientific mission operated on the international space
station (ISS). It carries two types of X-ray cameras: Gas Slit Camera
(GSC; \cite{2011arXiv1102.0891S}; \cite{2011arXiv1103.4224M}) and
Solid-state Slit Camera (SSC; \cite{2010PASJ...62.1371T};
\cite{2011arXiv1101.3651T}), covering the energy bands of 2--30 keV and
0.5--12 keV, respectively. MAXI/GSC observes nearly the whole sky
every 92 minutes with two instantaneous fields of view of $160^\circ
\times 3^\circ$. One of the main goals of the MAXI mission is to
provide a new all-sky X-ray source catalog, including both transient
and persistent objects. By integrating the data over a long period,
MAXI/GSC is expected to achieve so far the best sensitivities as an
all-sky mission that covers the 2--10 keV band (\cite{2009PASJ...61..999M}, 
\cite{2010fym..confE..35U}). Since its energy band is complementary to that of
the RASS (below 2 keV) and to those of Swift/BAT and INTEGRAL (above
10 keV), it will have an advantage in detecting sources having an
intrinsically soft continuum with moderate absorption.

In this paper, we present the first MAXI/GSC source catalog detected
in the 4--10 keV band at high Galactic latitudes ($|b| > 10^{\circ}$),
utilizing the first 7-month data since the start of its nominal
operation. The data reduction and filtering are given in section~2.
Section~3 describes the background model of MAXI/GSC used in the
analysis and the details of image analysis procedure. In section~4, we
present the source catalog and summarize the X-ray properties, results
of cross correlation with other catalogs and identification, position
accuracy, and log $N$ - log $S$ relations. Section~5 gives the
conclusion. The analysis of the local luminosity function of Seyfert
galaxies based on this catalog is reported in an accompanying paper
(Ueda et al.\ 2011, submitted to PASJ).

\section{DATA REDUCTION}
\label{sec:DATA_REDUCTION}

For the catalog production, we use the MAXI/GSC data taken between
2009 September 1 and 2010 March 31, when all the GSC counters were
operated with a high voltage of 1650 V. Only the data in the 4--10
keV band are utilized in this paper, 
because in this energy band (1) the energy and
position responses are best calibrated at present, and (2) a high
signal-to-noise ratio is achieved thanks to high detection efficiency
of the counters and to relatively low background rate (\cite{2011arXiv1102.0891S}).

Starting from the event files with processing version 0.3 provided by
the MAXI team, we apply the following data screening to obtain clean
data used in the image analysis described in section~\ref{sec:ANALYSIS}. 
The processed event files contain columns of the arrival time (TIME),
energy (PI), and sky position (R.A.\ and Dec.\ ) for each photon as
essential information. We utilize the data of all the twelve GSC
counters from 2009 September 1 to 22, while those of only eight
counters (GSC\_0, 1, 2, 3, 4, 5, 7, and 8) are included in the later
epoch since the operation of the rest four counters were stopped due to
a hardware trouble. The photon events detected by the carbon anodes \#1
and \#2 in all the counters are excluded because of the response
problem in the current calibration. To discard the data suffering from
high background rates, we only utilize the data taken when the ISS
latitude is between $-$40$^{\circ}$ and 40$^{\circ}$, and those detected
at the central part of each counter with the photon incident 
angle $|\phi| < 38^\circ$ ($\phi$; for definition, see \cite{2011arXiv1103.4224M}).

\begin{figure}
  \begin{center}
    \rotatebox{0}{
      \FigureFile(80mm,80mm){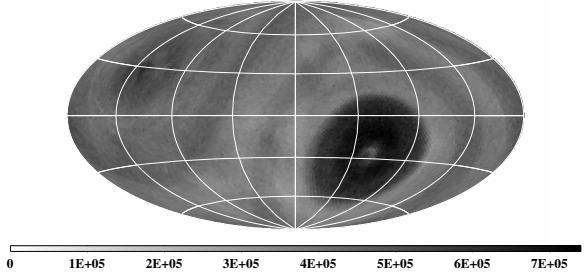}
    }
  \end{center}
  \caption{
The effective exposure map for the 7-month MAXI/GSC data in the
Galactic coordinates projected with the Aitoff algorithm. 
The unit is s cm$^2$.
The black annular-like structures at the bottom-right and top-left
positions correspond to regions of the longest exposure near the two poles
of the rotation axis of the ISS's orbital motion.  }
  \label{fig:exposuremap}
\end{figure}

Figure~\ref{fig:exposuremap} displays the effective exposure map in the Galactic
coordinates for the MAXI/GSC 7-month data, where the net exposure
corrected for the detection efficiency, multiplied by the projected
area of the slit ($\propto {\rm cos} \phi$), is given in units of s
cm$^2$ at each sky position. This is obtained from the simulation
utilizing the MAXI simulator {\bf maxisim} (\cite{2009aaxo.conf...44E}) 
by assuming a uniformly
extended emission as the input source; hence this plot is inevitably
smoothed by $\sim$3 degree, an angular resolution of MAXI.

\section{ANALYSIS}
\label{sec:ANALYSIS}

To detect X-ray sources in an unbiased way from the
MAXI/GSC data, we perform an image analysis where the image response,
background, and detailed observational conditions are taken into
account. We first examine the properties of the MAXI/GSC background on
the basis of the on-board data, and attempt to construct a background
model (section~\ref{subsec:Background_Reproduction}). 
Then, in section~\ref{subsec:Image_Analysis}, we employ a two-step approach
to search for source candidates, and determine their fluxes and
positions.

\subsection{Background Reproduction}
\label{subsec:Background_Reproduction}

In order to securely detect faint X-ray sources with a minimum number
of fake detections and to determine their X-ray fluxes precisely, it is
of crucial importance to reproduce the instrumental non-X-ray
background (NXB) level with a high accuracy. Thus, we model the
MAXI/GSC background as a function of time, on the basis of the
observational properties. In the case of MAXI, it is difficult to
disentangle the cosmic X-ray background (CXB) and NXB from the on-board
data, since MAXI rarely observes a direction of the night earth that
blocks the CXB. Therefore, we here consider the sum of these two
background components.

First, we examine the long-term variability of the background. For this
purpose, we make daily averaged background count rate for each counter by
rejecting periods when bright X-ray point sources or the bright region
along the Galactic plane with $|l| < \timeform{50\circ}$ and $|b| <
\timeform{10\circ}$ are within the GSC field of view. We find that the
background count rate is significantly different from counter to
counter. This is probably due to the difference in the configuration
of the counter relative to the ISS and to its direction of orbital
motion. Moreover, the background count rate is found to be highly
variable on a time scale of days, dependent on the various observational
conditions, such as the motion of the solar paddles and the shuttle
docking to the ISS.

\begin{figure}[htbp]
  \begin{center}
    \FigureFile(80mm,80mm){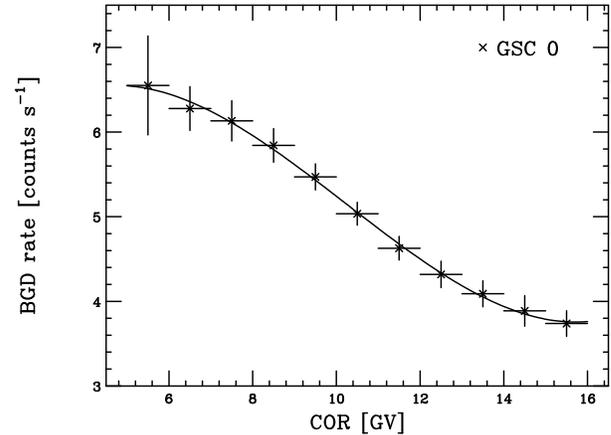}
  \end{center}
  \caption{
An example of COR dependence of the GSC background count rate 
for the counter GSC\_0. 
The peak of the COR-sorted daily-averaged background count rate
distribution is plotted, together with its standard deviation.  }
  \label{fig:COR_BGD}
\end{figure}

It is widely known that the short-term variation of the background is
tightly correlated with cut-off rigidity (COR) for an X-ray instrument on a low-earth
orbit (e.g., \cite{1989PASJ...41..373H,2008PASJ...60S..11T}). We
confirm such a trend also in the case of MAXI/GSC
(\cite{2011arXiv1102.0891S}), as shown in figure~\ref{fig:COR_BGD},
which plots the GSC background count rate against COR of the ISS
position. We model this background-COR relation by a third order
polynomial, which is shown with the solid line in figure~\ref{fig:COR_BGD}.

Utilizing the MAXI simulator (\cite{2009aaxo.conf...44E}), we create
background events for the individual counter, with a rate predicted by
the background-COR correlation given in figure~\ref{fig:COR_BGD}, 
after the long-term variation is
considered. The detector coordinate ${\rm DETX}$ and the pulse height
of each event are randomly assigned in the simulator according to the
observed distributions of ${\rm DETX}$ and pulse height, both of which
depend on COR (\cite{2011arXiv1102.0891S}). The simulated event list
is then processed in the same way as for the real data, and the sky
position in R.A.\ and DEC.\ are assigned to each event by referring to the
attitude of MAXI and the orbital motion of the ISS.

\subsection{Image Analysis}
\label{subsec:Image_Analysis}

Source detection analysis is performed for a tangentially projected
image of a small area in the ``sky coordinates'', as in the usual case
of data analysis of pointing satellites. An image in the sky
coordinates, $(X, Y)$, is defined by the reference point in R.A.\ and
Dec.\ corresponding to the image center, the pixel size in units of
degree in each direction, and the position angle between the $+Y$-axis
and the north direction, which is set to be zero in our analysis. The
entire sky is divided into 768 square images with a size of $14^{\circ}
\times 14^{\circ}$ so that neighboring regions have enough overlapping
area with a width of $>$3 degree in both sides, not to miss any sources
near the boundary. The center positions of the images have equal spacing
one another, which are adopted from those of the
``HEALPix''\footnote{http://healpix.jpl.nasa.gov} grid
(\cite{2005ApJ...622..759G}).  Each sky image is produced in the 4--10
keV band from the event list, referring to the columns of energy (PI)
and sky position (R.A.\ and Dec.\ ). The bin width is set to be 0.1
degree, sufficiently finer than the point spread function of MAXI (\cite{2011arXiv1102.0891S}).

\subsubsection{Step 1: Searching for Source Candidates}
\label{subsubsec:step1}

\begin{figure*}
  \begin{center}
    \rotatebox{0}{
      \FigureFile(80mm,60mm){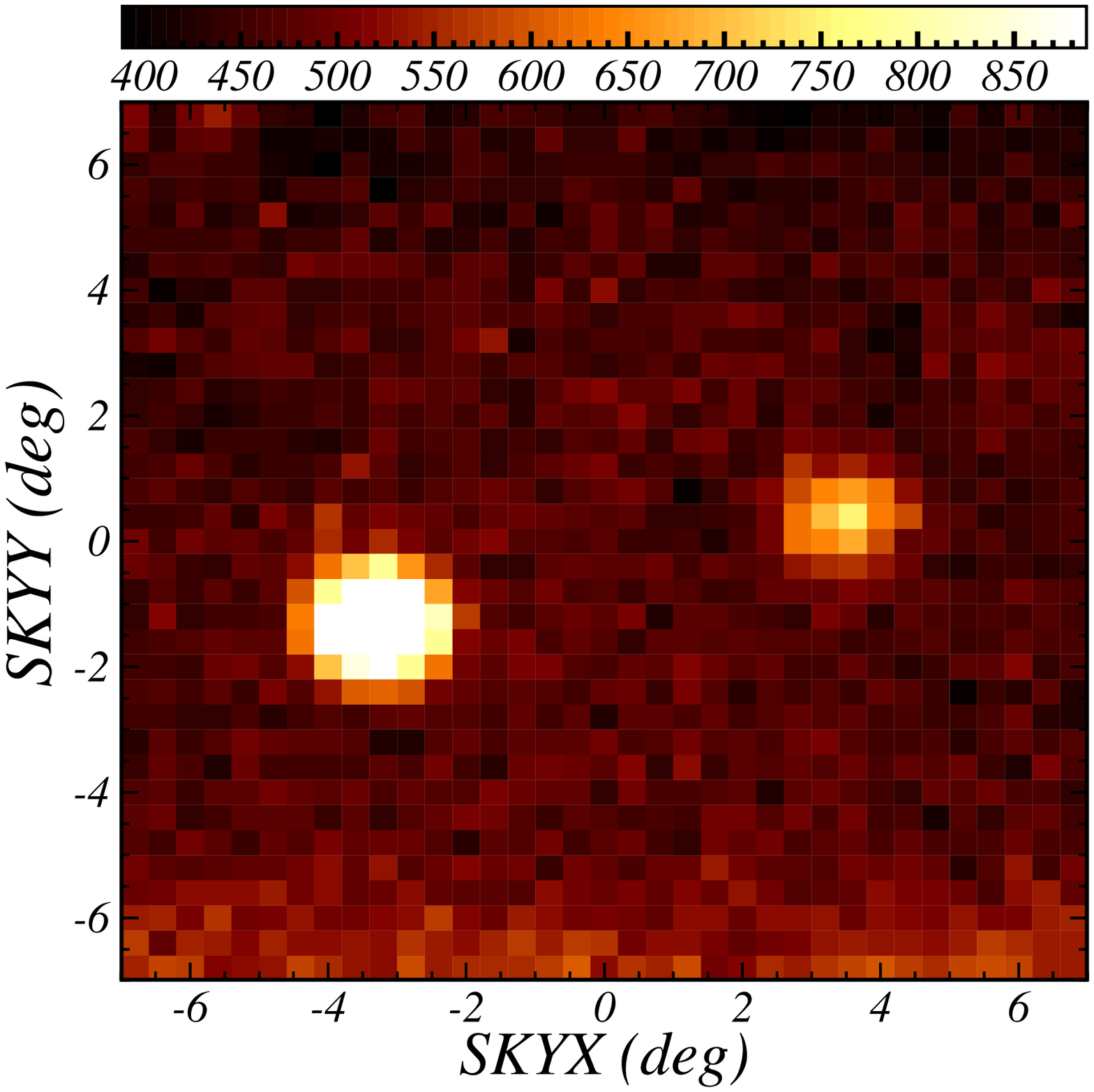}
      \FigureFile(80mm,60mm){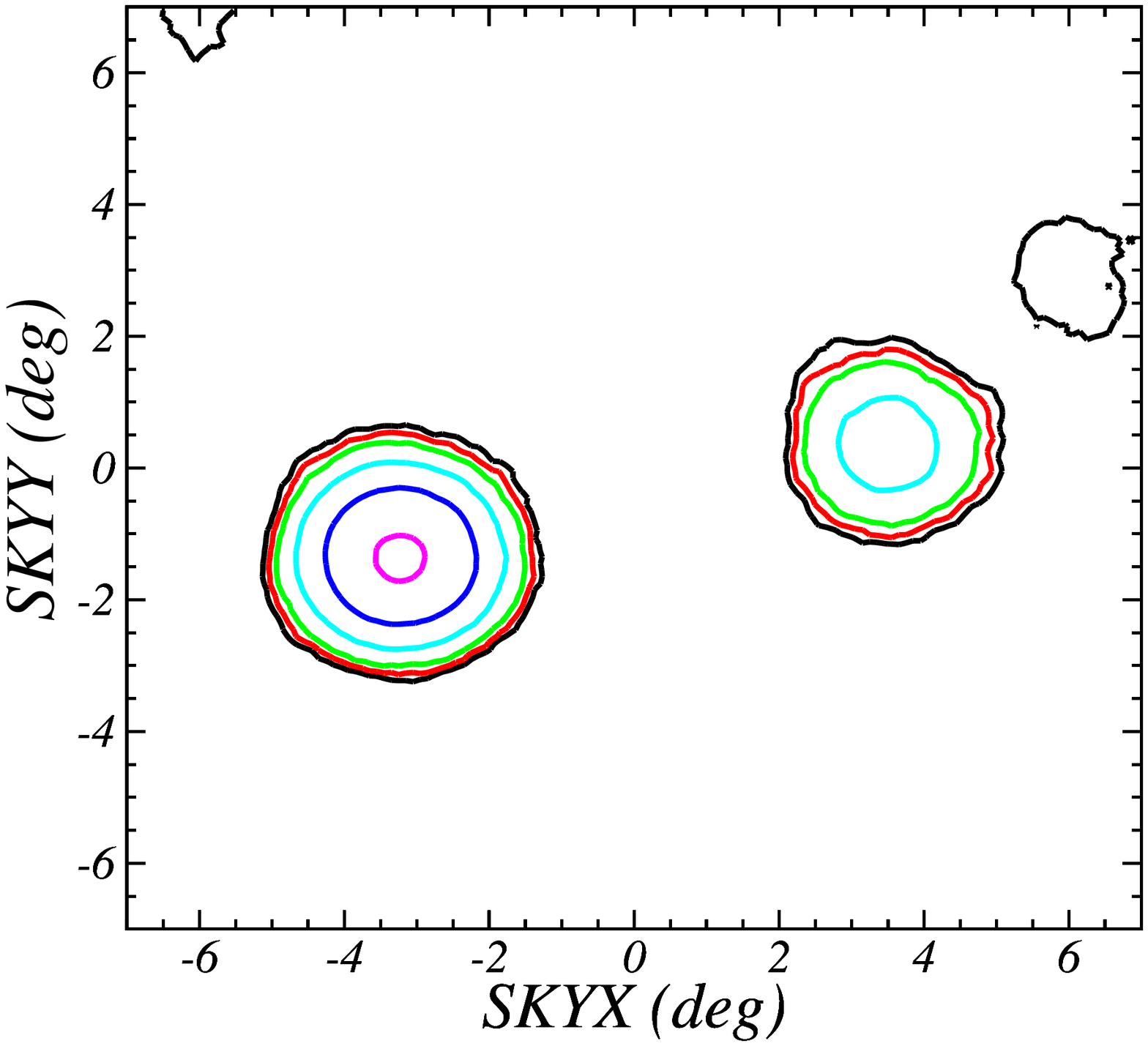}
    }
  \end{center}
  \caption{ (Left) An example of observed image in photon count space.
    (Right) The significance
    map of the same region. The black, red, green, cyan, blue,
    and purple contours represent the levels of significance of 5, 7,
    10, 20, 40, and 70, respectively.}
  \label{fig:rawdata_and_sigmap}
\end{figure*}

For a divided region of the entire sky as noted above, 
we prepare the images of the real data and simulated background data.
The background image is produced on the basis of 
the model described in section~\ref{subsec:Background_Reproduction}, 
by generating 10 times more photons
than the actual rate to make the statistical errors negligibly small
compared with those in the observed data. By subtracting the
background from the data, we obtain the ``residual'' image. An example 
of the real data is displayed in the left panel 
of figure~\ref{fig:rawdata_and_sigmap}.

To find source candidates from statistical argument, the data and
residual images are smoothed with a circle of $r = 1^{\circ}$ with a
constant weight of unity (i.e., simple integration). From the
integrated counts of the data and residual, we calculate the excess
significance at each point as ``$ {\rm residual} / \sqrt{{\rm data}}$'',
thus producing the ``significance map'' (right panel of
figure~\ref{fig:rawdata_and_sigmap}). Then, we search for a peak
showing the highest significance in the given image, which is listed
as a source candidate if it exceeds 4$\sigma$. Next, we mask the
circular region around its peak position with a radius of 3 degree,
where the signals are expected to originate from the same source due
to the point spread function. This step is repeated to
find another source, which is added to the list of source candidates
until the significance of the peak becomes lower than 4$\sigma$.

In this step, the normalization level of the background profile is
estimated from the data themselves for each image, to absorb any
remaining systematic errors in reproduction of the background rate,
and also to approximately take into account contamination of Galactic
diffuse emission in the background. To do this, we first detect bright
sources above 15$\sigma$ from the significance map produced by assuming
the nominal background rate predicted by the model. Then, after masking
out circular regions of $r=3^{\circ}$ around these bright sources, we tune
the normalization of the background level so that its total count 
matches that of the observed data in the same sky region. The tuned
level could be slightly overestimated because sources with
significances less than 15$\sigma$ are ignored in the first stage. This
is not a problem, however, since we set a conservative threshold for
picking up the source candidates ($4\sigma$) compared with that for
the final catalog (7$\sigma$). Furthermore, as described later
(section~\ref{subsubsec:step2}), we will make iteration of this process to ensure the
completeness of the source search by adopting a more accurate background
level.

After performing the source search in all the images, 
we merge the source
candidate lists from all of them. In some cases, one source can be
independently detected in multiple images, especially when it is
located around the edge of the images. We regard any pairs of source
candidates whose positional separation is smaller than 1.0 degree
are the same object, and we only leave the source detected at the 
closer position to the image center in the merged list. 
Further, we also eliminate the source candidates which are 
regarded as obviously fake detections by visual inspection, 
such as those detected around a very
bright source (like Sco X-1) or over strong extended diffuse emission
from a galaxy cluster. We finally have 499 source candidates above
4$\sigma$ at $|b| > 10^{\circ}$ in Step~1.

\subsubsection{Step 2: Determining Flux and Position}
\label{subsubsec:step2}

As preparation for the image fitting analysis, we construct the point spread
function (PSF) model of each source candidate identified in Step~1.  We
assume that all sources can be treated as point-like in our data. One
complexity in the analysis of the MAXI data is that the PSF depends on the
position of the detector, and hence its integrated shape in the sky
coordinates over multiple orbits is determined by the orbit and
attitude condition of MAXI (\cite{2011arXiv1102.0891S}). To take this
into account, we utilize the MAXI simulator
(\cite{2009aaxo.conf...44E}) to construct the PSF model in the sky
image under the exactly same conditions as for real data. To suppress
the statistical errors in the model, a sufficiently larger number of
photons are generated by adopting the Crab Nebula flux (1 Crab) for
all source candidates. In the simulation, we choose the same spectrum
as the Crab Nebula
\footnote{The spectral parameters are adopted from the
INTEGRAL General Reference Catalog
(ver. 31).~http://www.isdc.unige.ch/integral/science/catalogue}
: an absorbed power law with a photon index 
of 2.1 and a column density of 2.6$\times$10$^{21}$ cm$^{-2}$. 
We confirm that the choice of spectrum does not affect the flux determination,
even though the PSF also has a weak energy dependence. Comparing 
the PSF with 
the data of the brightest point source Sco X-1, we find that 
the PSF model reproduces the data with residuals corresponding to 
$\sim$4\% level of the peak height at $r \sim 1.0$ degree.
These systematic errors in the PSF profile do not affect the flux 
determination of any sources in our catalog that can
be regarded as point-like compared with the PSF size.

\begin{figure}
  \begin{center}
    \rotatebox{0}{
      \FigureFile(80mm,60mm){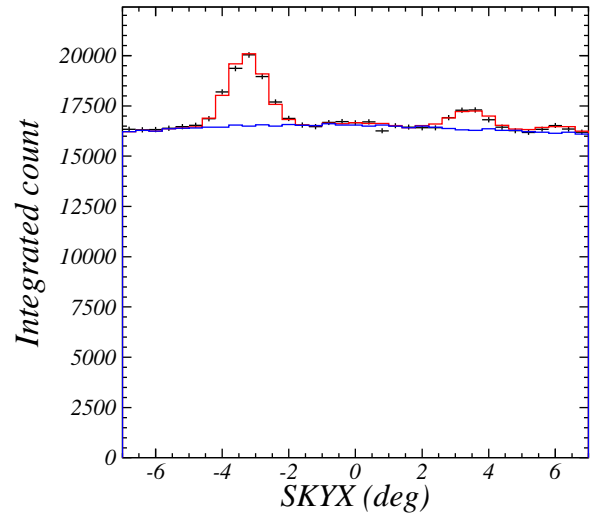}
    }
  \end{center}
  \caption{
    Comparison of the observed data (black) and the best-fit model
    (red: total, blue: only background). The faintest source in this
    image is located at $X=6.0$ with $s_{\rm D} \approx 7$.
  }
  \label{fig:fitimg}
\end{figure}

To determine the flux and position of the source candidates, we
perform image fit to the real data in photon count space with a model
consisting of the PSFs from the sources and the background. Only the
inner region of $11^\circ \times 11^\circ$ is utilized to ignore the
contribution from PSFs of sources located just outside the whole
image region of $14^\circ \times 14^\circ$.
To properly treat statistics with small numbers of photons in each
bin, the Poisson maximum likelihood algorithm with MINUIT package 
(\cite{1975CoPhC..10..343J}) is
adopted, based on the so-called $C$ statistics (\cite{1979ApJ...228..939C}), defined as 
\begin{equation}
    C \equiv 2 \sum_{i,j} \{M(i,j) - D(i,j) \ln{M(i,j)}\},
\end{equation}
where $D(i,j)$ and $M(i,j)$ represent the data and model at the image
pixel $(i,j)$, respectively. The best-fit parameters are obtained by
minimizing the $C$ value, and the 1$\sigma$ statistical error of a
single parameter can be estimated by finding that giving the $C$ value
larger by unity than the best-fit. In the fitting process, the
normalization of the PSF (i.e., flux) and its position are set to free
parameters for all the source candidates, as well as the background level. 
Figure~\ref{fig:fitimg}
shows the projection of the left panel of
figure~\ref{fig:rawdata_and_sigmap} onto X-axis (black), superposed
with the best-fit model (red: total, blue: background).
Since we treat all the sources as point-like, a few bright
galaxy clusters may have a small uncertainty in the flux due to their
extended structures significantly larger than the PSF (see Note in
table~1).

We define the detection significance ($s_{\rm D}$) of each source as
\begin{equation}
  s_{\rm D} \equiv (\mbox{best-fit flux}) / (\mbox{its}~1\sigma~\mbox{statistical error}).
\end{equation}
In this paper, we adopt $s_{\rm D} > 7$ as the detection criteria
for the final catalog, which is conservative enough not to contain any
fake detections within current systematic errors in the background
model. We verify this by checking the reproductivity of the background
profile on a scale of the PSF size ($\sim 3^{\circ}$) by the following
analysis. We search for ``negative'' signals in the residual image by
changing the sign of the significance map in Step~1. Then, we perform
image fitting with a model consisting of the background plus negative
PSFs identified in the previous step. As a result, it is confirmed that
no negative peaks are detected with significances above 7$\sigma$ from
all the images. This indicates that the number of fake, positive peaks
is also expected to be zero.

Using the tentative source list produced in this way, we repeat the
same procedure of Step~1 by applying the best-fit ``background plus
PSFs'' model determined in Step~2, instead of the pure background
model as done in the first analysis. The main purpose is to search for
missing sources located close to a nearby, brighter source because of
the mask around the peak with a radius of 3 degree applied to the
significance map in Step~1. Finally, we perform image fitting by
adding these newly identified source candidates into the model. The final
number of the detected sources with $s_{\rm D}>7$ is 143 at
$|b|>10^\circ$, including additional 4 sources found by this
iteration process.

The solar paddle structures on the ISS sometimes block a part of the
field of view of MAXI, which works to reduce the apparent averaged
count rate of an X-ray source. Since we do not take into account these
effects in the image analysis, we must evaluate the shielding effect
to derive the correct fluxes of the cataloged sources. To estimate it,
we perform the whole sky simulations of the CXB with and without the
solar paddle occultation during the 7-month period. By comparing
these results, we estimate the fraction of  unocculted observing time
at each sky position. It is found to be $\approx$96\% on average and
in a range of 90\%--100\%, depending on the position. Thus, we
correct the flux obtained in Step~2 for this fraction according to the
sky position, which is finally listed in the catalog.

\section{MAXI CATALOG}
\label{sec:MAXI_CATALOG}

\subsection{X-Ray Properties}
\label{subsec:X-ray_Properties}

In table~1, we present the first MAXI/GSC catalog at high Galactic
latitudes ($|b| > 10^{\circ}$) compiled from the 7-month data. It
contains 143 sources detected with significance above 7$s_{\rm
D}$ in the 4--10 keV band, sorted by R.A.\ and Dec.\ . The 1st to 7th
columns give (1) source identification number, (2) MAXI source name
designated from the detected position, (3, 4) MAXI best-fit position in
R.A.\ and Dec., (5) detection significance $s_{\rm D}$, (6) flux
in the 4--10 keV band converted from the count rate by assuming a
photon index of 2.1, and (7) its $1\sigma$ statistical error.

%
%

\begin{figure*}
  \begin{center}
    \rotatebox{0}{
      \FigureFile(60mm,60mm){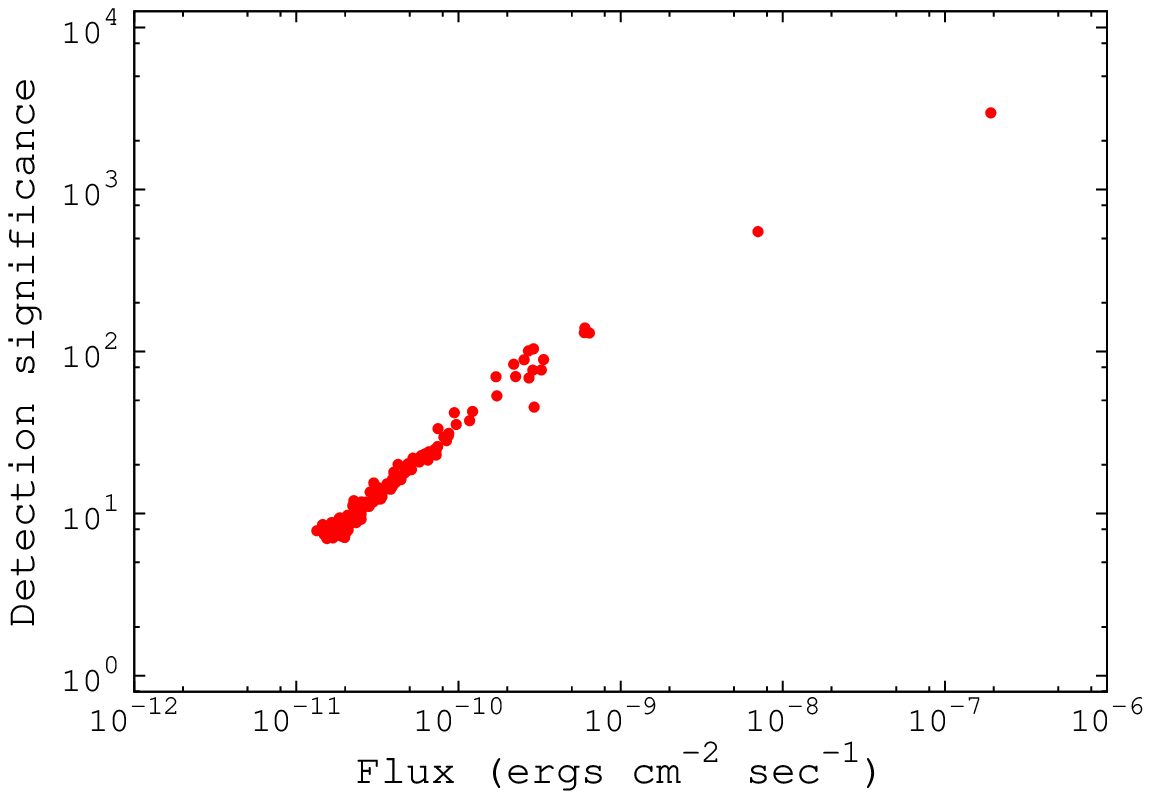}
      \FigureFile(60mm,60mm){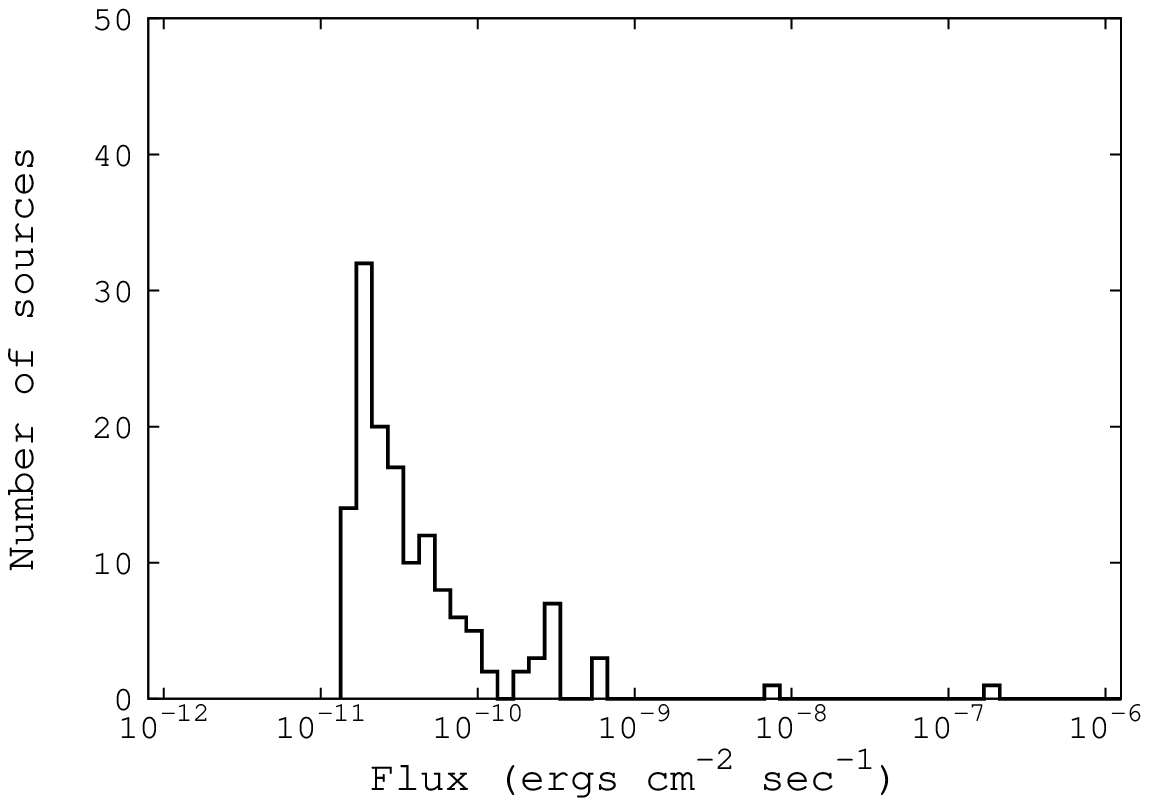}
      \FigureFile(60mm,60mm){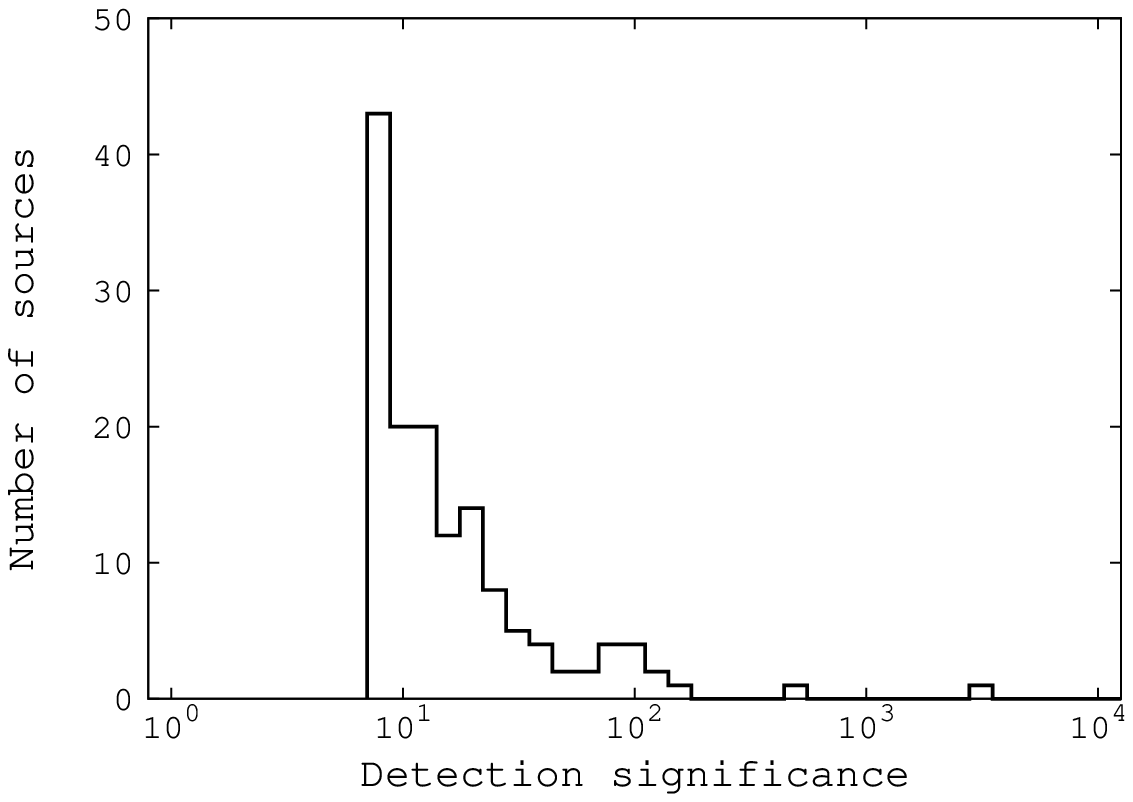}
    }
  \end{center}
  \caption{
    The correlation between the detection significance and the flux for all
    the sources in the MAXI/GSC catalog (left), the distribution of the flux
    (middle), and that of the detection significance (right).  }
  \label{fig:flux_sig}
\end{figure*}

The left panel of figure~\ref{fig:flux_sig} shows the correlation between
the detection significance ($s_{\rm D}$) and the flux. 
It is seen that  $s_{\rm D}$ is roughly proportional to the flux 
in the low flux region where the background determines the noise 
(statistical fluctuation of photon counts). Whereas in the high flux region, 
it is proportional to (flux)$^{1/2}$ since the source photons dominate the
noise.
The scatter is mainly due to the variation of exposure (see figure~\ref{fig:exposuremap}).
The histograms of flux and $s_{\rm D}$ are shown in the middle
and right panels of figure~\ref{fig:flux_sig}, respectively. 
The limiting sensitivity is found to be $\sim$1.2 mCrab with $s_{\rm D} >$ 7,
corresponding to $\sim1.5\times10^{-11}$ ergs cm$^{-2}$ s$^{-1}$ in
the 4--10 keV band.  This is fully consistent with our expectation
from the actual background level and observing efficiency of MAXI
(\cite{2010fym..confE..35U}).

\subsection{Identification}
\label{subsec:Identification}

Source identification is one of the most important task for the
catalog obtained from an unbiased survey. Since the position error of
MAXI is typically $0.4^{\circ}$ for sources with $s_{\rm D}=7$ at 90\%
confidence level (section~\ref{subsec:Position_Accuracy}), direct identification in the optical or
near infrared band solely based on the position is practically
difficult without further localization. Therefore, we cross correlate
our catalog with major X-ray/gamma-ray catalogs for which optical
identification or position determination with an $\sim$arcsec accuracy
is available, assuming that the MAXI sources are likely detected also
in these catalogs with similar flux limits.
Table~\ref{tab:ref_catalog} lists the catalogs and references we use
for the cross correlation; Swift/BAT 58-month and 
54-month Catalogs, 1st Fermi/LAT AGN Catalog, 
the brightest 2200 sources in the 0.5--2 keV band 
in the RASS BSC, INTEGRAL General Reference Catalog (version 31), 
NORAS Galaxy Cluster Survey Catalog, REFLEX Galaxy Cluster Survey
Catalog, and 1st XMM-Newton Slew Survey Catalog.

%
%
\setcounter{table}{1}
\tiny
\input{ref_catalog.tex}
\normalsize

To cross correlate the MAXI/GSC catalog with the reference ones, we
only take into account the position error of MAXI, $\sigma_{\rm
pos}$, since those in the other catalogs are negligibly small. It
is defined as
\begin{equation}
  \sigma_{\rm pos} \equiv \sqrt{\sigma^2_{\rm stat} + \sigma^2_{\rm sys}}, 
\end{equation}
where $\sigma_{\rm sys}$ is the systematic error of 0.05 degree
(1$\sigma$, section~\ref{subsec:Position_Accuracy}) and $\sigma_{\rm stat}$ is 
the statistical one combined
from those in the $X$ and $Y$ directions at 1$\sigma$ obtained in the image fit
process (Step~2). To be conservative, we adopt $3\sigma_{\rm pos}$ as the
radius for the position matching. We find that this roughly
corresponds to a confidence level of 99\% in the 2-dimensional
space, giving a typical increment in the $C$ value by 9.2.

The total numbers of possible counterparts found from each reference
catalog are also summarized in table~\ref{tab:ref_catalog}. 
We identify the MAXI sources
largely on the basis of this cross-correlation results. Basically, we first
look for counterparts in the Swift/BAT and INTEGRAL catalogs, which
achieve similar sensitivities to that of MAXI for sources without heavy
obscuration. If no sources are found from these catalogs, we also
refer to the results with the other catalogs. Ten MAXI sources are
found to have no matched counterparts in any of these catalogs within 3$\sigma_{\rm pos}$. 
In such cases, we identify the counterpart by individual inspection using
NED\footnote{http://ned.ipac.caltech.edu} and 
SIMBAD\footnote{http://simbad.u-strasbg.fr/simbad}.

To estimate the number of possible spurious
identification, we calculate the expected number of coincidental
matches between the MAXI sources and those in the reference catalogs,
$N_{\rm cm}$, using the following equation:
\begin{equation}
  N_{\rm cm} = \sum_{i}\rho_{i} \times S_{i},
\end{equation}
where the suffix $i$ denotes each reference catalog that can be
considered to be independent, and $\rho_{i}$ and $S_{i}$ are the mean
surface number-density of sources in the reference catalog and the
total area within the $3\sigma_{\rm pos}$ error radii of MAXI sources
searched for their counterparts in that catalog, respectively. Here we
assume that the spatial distribution of sources in each catalog can be
regarded to be random, which is a good approximation in the high
Galactic-latitude sky. For simplicity, we only refer to the Swift/BAT
58-month Catalog and the RASS BSC for this calculation, which cover
the different energy bands and hence are complementary each
other. Following our actual identification procedure, we first
consider the Swift/BAT 58-month Catalog and the error region of all
MAXI sources, which correspond to $\rho_1 = 0.022$ deg$^{-2}$ and $S_1
= 42$ deg$^2$, respectively. Then, to reflect the fact that the
remaining MAXI sources without hard X-ray counterparts have been
mainly identified by the RASS catalog, we adopt $\rho_2 = 0.054$
deg$^{-2}$ for the 1829 brightest RASS sources and $S_2 = 18$ deg$^2$
for the MAXI error regions. In this way, we obtain $N_{\rm cm}$ =
1.9. Note that this is a conservative estimate since we have neglected
the duplicative sources between the Swift/BAT and RASS catalogs in the
calculation.

\begin{figure*}
  \begin{center}
    \rotatebox{0}{
      \FigureFile(160mm,90mm){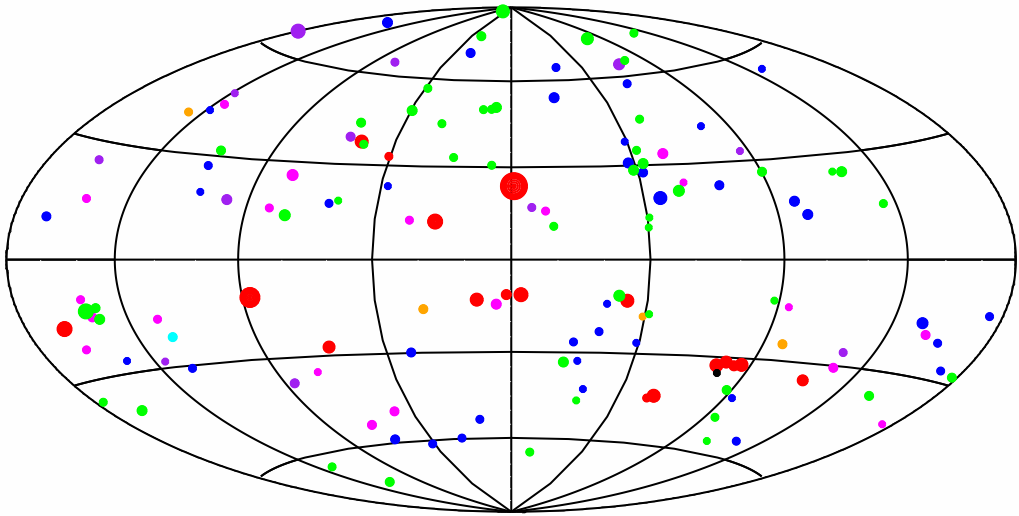}
    }
  \end{center}
  \caption{
    Locations of cataloged sources with $s_{\rm D} >$ 7 in 
    the Galactic coordinates. The radius is proportional to 
    logarithm of the 4--10 keV flux. Different colors correspond to  
    different types of sources: unidentified (black); galaxies (cyan); 
    galaxy clusters (green); Seyfert galaxies (blue); blazars (purple); 
    CVs/Stars (magenta); X-ray binaries (red); and confused (orange).
  }
  \label{fig:allsky_map}
\end{figure*}

Out of the total 143 MAXI sources, we finally identify 142 sources,
and only one source\footnote{MAXI J0457$-$696} remains unidentified. The
locations of the cataloged sources are plotted in figure~\ref{fig:allsky_map} in the
Galactic coordinates with different colors corresponding to different
types of object. Table~\ref{tab:category} gives a summary of source
identification. The catalog contains 38 Galactic/LMC/SMC objects, 48
galaxy clusters, 51 AGNs including 12 blazars. The detailed
information of the counterparts is provided in the 8th to 14th columns
of table~1 for each MAXI source, (8) the source name, (9, 10) the
position of the counterpart in R.A.\ and Dec., (11) type, (12) redshift
(only for extragalactic objects), (13) other names of the source, and
(14) notes in special cases. We find four MAXI sources\footnote{MAXI
J0627$-$540, J0957+693, J1633$-$750, and J1941$-$104} are probably 
confused from multiple objects
that are difficult to be resolved with the angular resolution of MAXI.
For those sources, multiple counterparts are listed in table~1, 
which are not counted in the identification summary shown in table~\ref{tab:category}.

We compare the number statistics of source populations in the first
MAXI/GSC catalog with those in the HEAO-1 A-2 ($|b|>20^\circ$,
\cite{1982ApJ...253..485P}), RXTE ($|b|>10^\circ$,
\cite{2004A&A...418..927R}), and Swift/BAT
(\cite{2010A&A...524A..64C}) catalogs. 
The number ratios between Galactic objects, AGNs, and galaxy
clusters in our catalog, 21:40:36 ($|b|>20^\circ$)
and 38:51:48 ($|b|>10^\circ$), are found to be
consistent with both HEAO-1 A-2 (17:29:30) and RXTE (63:100:64)
results within the statistical errors. By contrast, the
Swift/BAT survey performed in the 15--150 keV band
brings a significantly higher fraction of AGNs at $|b| > 10^\circ$,
$\sim$60\%, than these catalogs produced in lower energy bands.
We confirm that
all but two galaxy clusters in the HEAO-1 A-2 catalog are listed in
ours, with the remaining two also detected with slightly lower
significances than $s_{\rm D} = 7$ in the MAXI/GSC data. 
We note, however, that about 40\% (= 12/29) of the Piccinotti AGNs are
absent in our catalog. In fact, the X-ray
fluxes of these 12 AGNs obtained from pointing observations with ASCA,
XMM-Newton, or Swift/XRT in more recent years
(\cite{2009ApJ...690.1322W}, \cite{2006AJ....131.2843S},
\cite{2005A&A...442..895A}, and \cite{2004A&A...417...61B}) are all
lower than the typical limiting sensitivity of the MAXI survey,
$\approx$2$\times$10$^{-11}$ ergs cm$^{-2}$ s$^{-1}$ in the 4--10 keV
band (figure~\ref{fig:area}, in section
\ref{subsec:logN_logS_Relation}), by assuming a power law photon index
of 2. This fact indicates that the list of the
brightest X-ray AGNs in the whole sky has significantly changed since
$\sim$30 years ago due to their long-term variability.

\scriptsize
\input{type-table.tex}
\normalsize

\subsection{Position Accuracy}
\label{subsec:Position_Accuracy}

\begin{figure*}
  \begin{center}
    \rotatebox{0}{
      \FigureFile(50mm,50mm){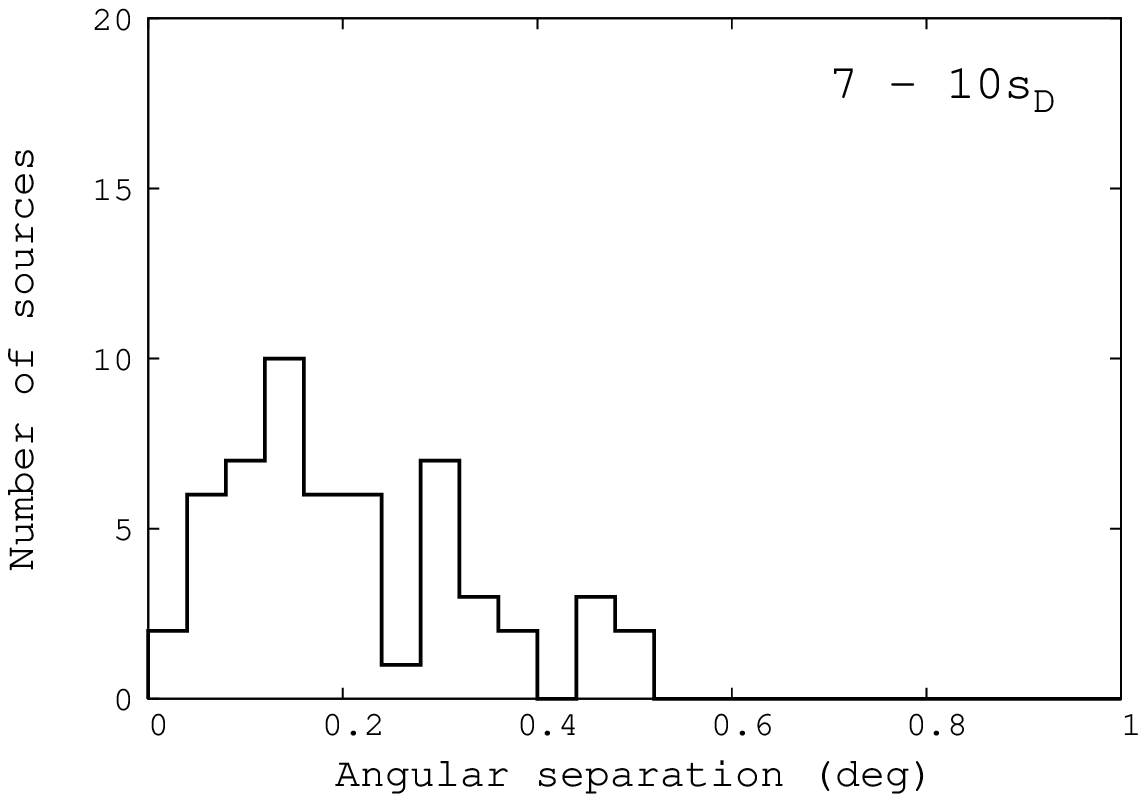}\hspace{5mm}
      \FigureFile(50mm,50mm){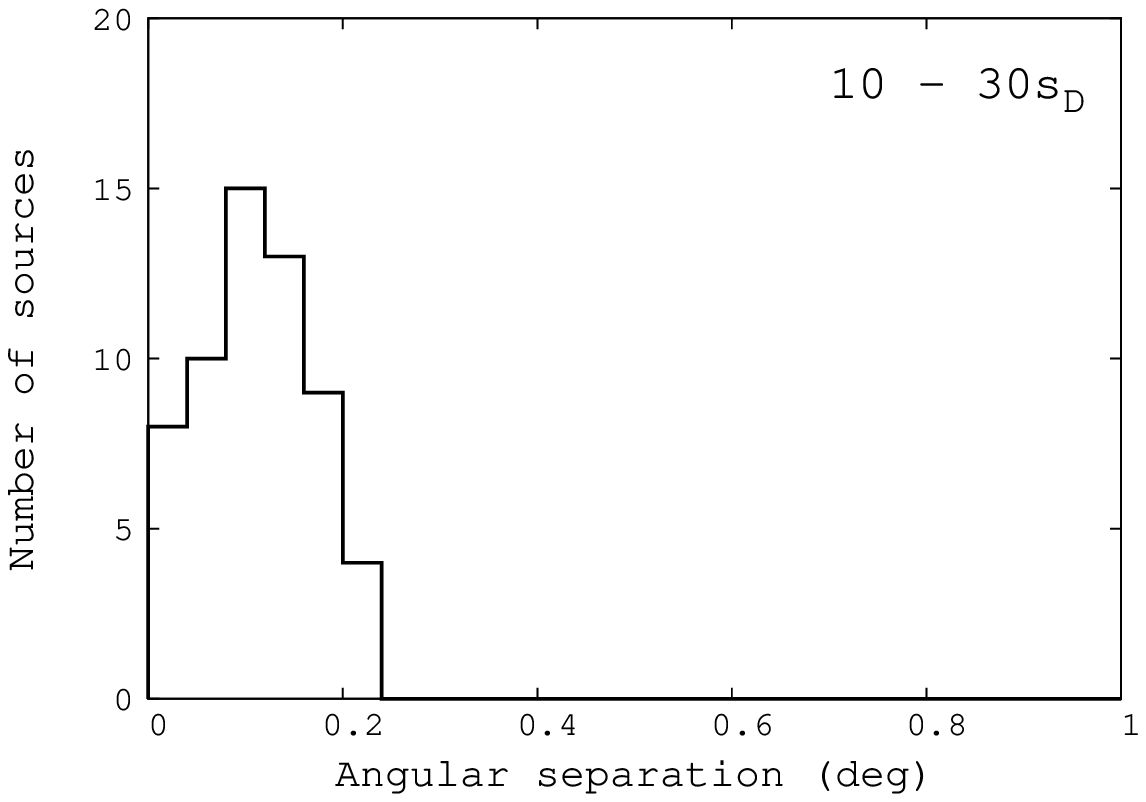}\hspace{5mm}
      \FigureFile(50mm,50mm){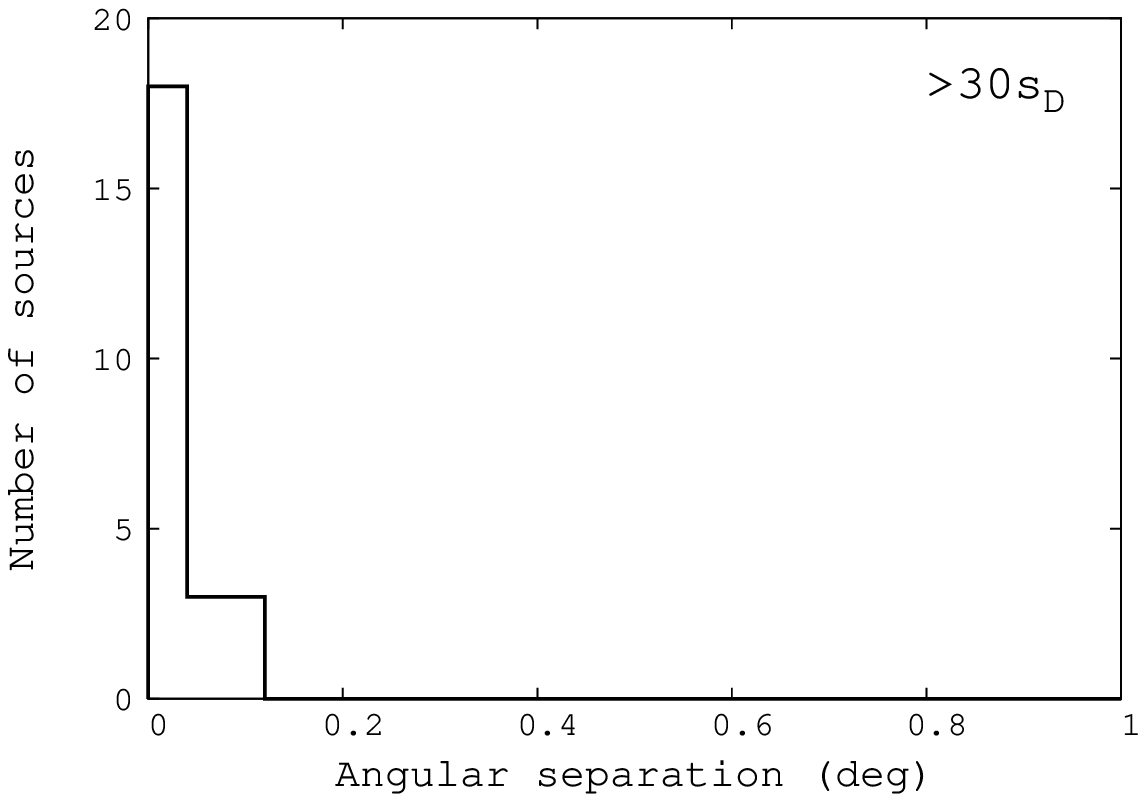}
    }
  \end{center}
  \caption{
    Source distributions as a function of the angular separation between the MAXI position and that of 
    the counterpart for different regions of detection significance (from left to right, 
    7--10$s_{\rm D}$, 10--30$s_{\rm D}$, and $>$30$s_{\rm D}$). 
  }
  \label{fig:position}
\end{figure*}

Utilizing the source identification result, we estimate the positional
error of MAXI as a function of the detection significance, which is
observationally determined. Therefore, it gives the most robust information
useful for identification work of MAXI sources within the
current calibration. Figure~\ref{fig:position} shows the source distributions 
as a function of the angular separation between the MAXI position and that of the
counterpart in different $s_{\rm D}$ ranges; $s_{\rm
D}=7-10$ (left), $s_{\rm D}=10-30$ (middle), and $s_{\rm
D}>30$ (right). From each distribution, we measure a 90\% error
radius by counting the number of sources.

\begin{figure}
  \begin{center}
    \rotatebox{0}{
      \FigureFile(80mm,60mm){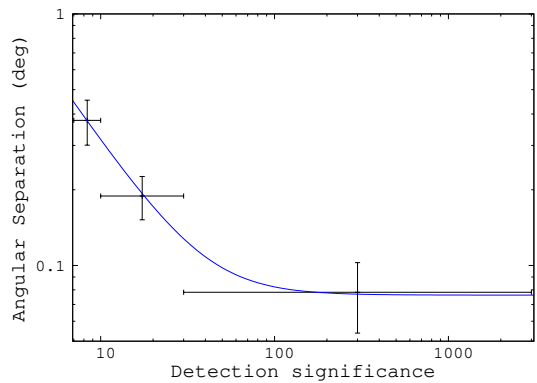}
    }
  \end{center}
  \caption{
    90\% error radius of MAXI sources plotted as a function of the 
    detection significance. The blue curve indicates the best-fit formula
    with the form of $\sqrt{(3.08/s_{\rm D})^2 + (0.08)^2}$ (degree).
  }
  \label{fig:angsep_significance}
\end{figure}

Figure~\ref{fig:angsep_significance} plots the 90\% error radius as a
function of the detection significance. In the highest $s_{\rm D}$
bin, the statistical errors are expected to be negligibly small compared
with the systematic errors due to those in the attitude determination
and position calibration of the GSC (\cite{2011arXiv1102.0891S}). 
We thus estimate the systematic error to be $\sigma_{\rm sys}^{90} \sim 0.08$ degree 
(90\% confidence level), or $\sigma_{\rm sys} \sim 0.05$ degree (1$\sigma$).
In the lower significance region, the total error is dominated by the statistical error, 
which is expected to be proportional to $s_{\rm D}^{-1}$. We obtain 
the best-fit formula for the 90\% positional error of MAXI sources 
\begin{equation}
  \sigma_{\rm pos}^{90} (s_{\rm D}) =  \sqrt{(A/s_{\rm D})^2 + (\sigma_{\rm sys}^{90})^2}, 
\end{equation}
where $A=3.08\pm0.04$ and $\sigma_{\rm sys}^{90} = 0.08$.

\subsection{log $N$ - log $S$ Relation}
\label{subsec:logN_logS_Relation}

\begin{figure}
  \begin{center}
    \rotatebox{0}{
      \FigureFile(80mm,60mm){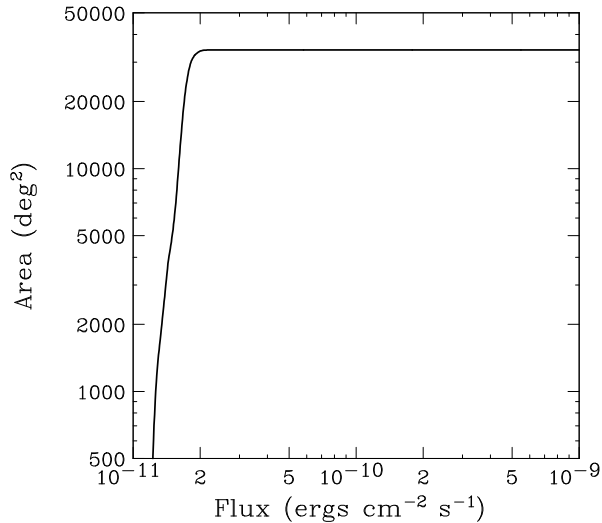}
    }
  \end{center}
  \caption{
    Area curve of the 7-month MAXI/GSC survey at $|b|>10^{\circ}$ in the
    4--10 keV band, for the detection significance of $s_{\rm D} > 7$.
}
  \label{fig:area}
\end{figure}

Source number counts (log $N$ - log $S$ relation) give the most
fundamental statistical properties of source populations that can be
derived from survey observations. To obtain this, we need to have an
area curve, where survey area guaranteed for detection of a source
with the given detection criteria (i.e., $s_{\rm D} >7$) is given as a
function of flux. We estimate the sensitivities at each sky position
from (1) the background photon counts and (2) the product of effective
detector area $\times$ exposure obtained from the CXB simulation
(shown in figure~\ref{fig:exposuremap}). 
We confirm that the actual detection significance
derived from the PSF fit can be well approximated by that analytically
calculated from these parameters on the basis of a simple statistical
argument. Figure~\ref{fig:area} plots the area curve calculated in
this way; we confirm that the $7\sigma$ sensitivity is $\approx$1.2
mCrab, or $1.5\times10^{-11}$ ergs cm$^{-2}$ s$^{-1}$, in the 4--10 keV band.

\begin{figure}
  \begin{center}
    \rotatebox{0}{
      \FigureFile(80mm,60mm){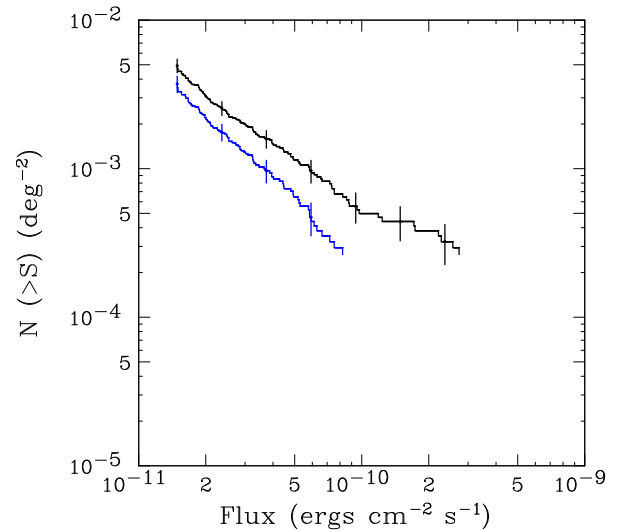}
    }
  \end{center}
  \caption{ 
    Log $N$ - log $S$ relations in the 4--10 keV band obtained from
    the 7-month MAXI/GSC survey at $|b|>10^{\circ}$. From top to bottom,
    those of the total sources (black) and extragalactic objects
    (blue). The error bars correspond to 
    the 90\% statistical errors in the source counts.
  }
  \label{fig:logN_logS}
\end{figure}

Dividing the flux distribution of the detected sources by the survey
area gives log $N$ - log $S$ relation in the differential
form. Figure~\ref{fig:logN_logS} shows that in the integral form, where the source
number density $N$ above flux $S$ is plotted. We separately plot
those of all the sources and extragalactic objects. 
The results for extragalactic objects are in an
excellent agreement with the HEAO-1 A-2 result by \cite{1982ApJ...253..485P},
by converting the fluxes from the 2--10 keV band into the 4--10 keV
band assuming a photon index of 2. 
We recall, however, that the actual content of the AGN sample has been
significantly changed since the HEAO-1 A2 era, as mentioned in section~\ref{subsec:Identification}.

\section{CONCLUSION}
\label{CONCLUSION}

We present the first MAXI/GSC catalog produced from an unbiased
X-ray survey in the 4--10 keV band at high Galactic latitude sky
($|b|>10^{\circ}$). The initial 7-month data after the start of normal
operation are utilized here. The limiting sensitivity of
$1.5\times10^{-11}$ ergs cm$^{-2}$ s$^{-1}$ (1.2 mCrab) is achieved
for the detection significance criteria of $>7 s_{\rm D}$, which is
conservatively adopted in this paper not to contain fake sources under
the current calibration. The sensitivity already exceeds that of the
HEAO-1 A-2 all-sky survey in the 2--10 keV band. It is expected to be
improved significantly by adding more data and by further improvement
of the background model calibration.

The catalog contains 143 sources, which are identified as 38
Galactic/LMC/SMC objects, 48 galaxy clusters, 39 Seyferts, 12 blazars,
1 galaxy, and 4 confused objects. Only one source remains
unidentified. The high completeness of this catalog makes it
particularly useful to investigate the statistical properties of X-ray
populations in the local universe with the least uncertainties. The
initial results on the X-ray luminosity function of Seyfert galaxies are
reported in the accompanying paper.

\bigskip

This research has made use of the NASA/IPAC Extragalactic Database (NED) 
which is operated by the Jet Propulsion Laboratory, California Institute 
of Technology, under contract with the National Aeronautics and Space Administration. 
This research has made use of the SIMBAD database, operated at CDS, Strasbourg, France.
Some of the results in this paper have been derived using the HEALPix 
(K.M. G{\'o}rski et al., 2005, ApJ, 622, p759) package.
This research was partially supported by the Ministry of Education, 
Culture, Sports, Science and Technology (MEXT), Grant-in-Aid No.19047001, 
20041008, 20244015 , 20540237, 21340043, 21740140, 22740120, 
23000004, 23540265, and Global-COE from MEXT ``The Next Generation of
Physics, Spun from Universality and Emergence'' and ``Nanoscience and
Quantum Physics''.

\end{document}


\begin{landscape}

\tabletypesize{\tiny}
\setlength{\tabcolsep}{0.035in}
\renewcommand{\arraystretch}{1.05}
\begin{deluxetable}{rlrrrrrlrrcllc}
\tablecaption{Cataloged sources in the MAXI/GSC survey.}
\tablewidth{0pt}
\tablehead{

  \multicolumn{1}{c}{(1)} & \multicolumn{1}{c}{(2)} & \multicolumn{1}{c}{(3)} & \multicolumn{1}{c}{(4)} & \multicolumn{1}{c}{(5)} & \multicolumn{1}{c}{(6)} & \multicolumn{1}{c}{(7)} & \multicolumn{1}{c}{(8)} & \multicolumn{1}{c}{(9)} & \multicolumn{1}{c}{(10)} & \multicolumn{1}{c}{(11)} & \multicolumn{1}{c}{(12)} & \multicolumn{1}{c}{(13)} & \multicolumn{1}{c}{(14)} \\ 
  \multicolumn{1}{c}{No.} & \multicolumn{1}{c}{MAXI Name} & \multicolumn{1}{c}{MAXI R.A.} & \multicolumn{1}{c}{MAXI Decl.} & \multicolumn{1}{c}{Significance} & \multicolumn{1}{c}{Flux\tablenotemark{a}} & \multicolumn{1}{c}{Error\tablenotemark{a}} & \multicolumn{1}{c}{Counterpart Name} & \multicolumn{1}{c}{Ctpt. R.A.} & \multicolumn{1}{c}{Ctpt. Decl.} & \multicolumn{1}{c}{Type\tablenotemark{b}} & \multicolumn{1}{c}{Redshift\tablenotemark{c}} & \multicolumn{1}{c}{Other Name} & \multicolumn{1}{c}{Note\tablenotemark{d}}

}
\startdata

1	&	MAXI J0042$-$093	&	10.5	&	$-$9.4	&	16.2	&	4.4	&	0.3	&	Abell 85                      	&	10.408	&	$-$9.342	&	Galaxy Cluster       	&	0.05506	&	4U 0037$-$10	&		\\
2	&	MAXI J0043+410	&	10.8	&	41.1	&	16.0	&	3.9	&	0.2	&	M31	&	10.678	&	41.179	&	Galaxy	&	 	&	Andromeda Galaxy	&		\\
3	&	MAXI J0048+320	&	12.1	&	32.0	&	10.4	&	2.4	&	0.2	&	Mrk 348                       	&	12.196	&	31.957	&	Sy2                  	&	0.01503	&	2E 0046.0+3140	&		\\
4	&	MAXI J0053$-$726	&	13.4	&	$-$72.7	&	15.5	&	3.0	&	0.2	&	2E 0050.1$-$7247	&	12.972	&	$-$72.530	&	HMXB	&	 	&		&		\\
5	&	MAXI J0055+462	&	13.8	&	46.3	&	8.3	&	1.9	&	0.2	&	1RXS J005528.0+461143         	&	13.867	&	46.195	&	CV/DQ Her	&	 	&	SWIFT J0055.4+4612	&		\\
6	&	MAXI J0056$-$012	&	14.2	&	$-$1.3	&	8.6	&	2.1	&	0.2	&	Abell 119	&	14.089	&	$-$1.263	&	Galaxy Cluster	&	0.0442	&	4U 0050$-$01	&		\\
7	&	MAXI J0117$-$734	&	19.3	&	$-$73.4	&	103.9	&	29.0	&	0.3	&	SMC X-1                       	&	19.271	&	$-$73.443	&	HMXB/NS              	&	 	&	3A 0116$-$736, 4U 0115$-$73	&		\\
8	&	MAXI J0125+341	&	21.3	&	34.1	&	7.0	&	1.5	&	0.2	&	1ES 0120+340	&	20.786	&	34.347	&	BL Lac	&	0.272	&	1ES 0120+34.0	&	\\
9	&	MAXI J0229+315	&	37.4	&	31.6	&	8.1	&	1.8	&	0.2	&	NGC 931                       	&	37.060	&	31.312	&	Sy1.5                	&	0.01665	&	1ES 0225+31.0	&		\\
10	&	MAXI J0254+416	&	43.7	&	41.6	&	22.8	&	5.9	&	0.2	&	WBL 088	&	43.634	&	41.586	&	Galaxy Cluster	&	0.017239	&	4U 0253+41	&	\\
11	&	MAXI J0258+133	&	44.7	&	13.4	&	19.6	&	4.9	&	0.2	&	Abell 401                     	&	44.737	&	13.582	&	Galaxy Cluster       	&	0.0748	&	2E 0256.2+1322	&		\\
12	&	MAXI J0301+443	&	45.3	&	44.3	&	13.1	&	3.3	&	0.2	&	CIZA J0300.7+4427	&	45.188	&	44.463	&	Galaxy Cluster	&	0.03	&	XSS J03008+4429	&		\\
13	&	MAXI J0308+409	&	47.0	&	41.0	&	17.7	&	4.6	&	0.2	&	Algol	&	47.042	&	40.958	&	Algol Type Eclipsing binary	&	 	&	2E 0304.9+4045	&  (A)	\\
14	&	MAXI J0317$-$443	&	49.3	&	$-$44.4	&	7.2	&	1.9	&	0.3	&	Abell 3112	&	49.494	&	$-$44.239	&	Galaxy Cluster	&	0.075252	&	1ES 0316$-$44.4	&		\\
15	&	MAXI J0319+415	&	50.0	&	41.5	&	130.9	&	59.7	&	0.4	&	Perseus Cluster	&	49.951	&	41.512	&	Galaxy Cluster	&	0.01756	&	4U 0316+41	& (B)	\\
16	&	MAXI J0327+288	&	51.8	&	28.8	&	8.7	&	1.9	&	0.2	&	UX Ari	&	51.646	&	28.717	&	RS CVn	&	 	&	3A 0322+277	&	(C)	\\
17	&	MAXI J0331+437	&	52.8	&	43.8	&	9.4	&	2.3	&	0.2	&	GK Per                        	&	52.799	&	43.905	&	CV/DQ Her            	&	 	&	3A 0327+438	&		\\
18	&	MAXI J0333$-$364	&	53.3	&	$-$36.4	&	7.8	&	2.0	&	0.2	&	NGC 1365                      	&	53.402	&	$-$36.140	&	Sy1.8                	&	0.00546	&	2E 0331.7$-$3618	&	\\
19	&	MAXI J0337+004	&	54.3	&	0.5	&	7.9	&	1.9	&	0.2	&	HR  1099	&	54.197	&	0.588	&	RS CVn	&	 	&	4U 0336+01	&	(D)	\\
20	&	MAXI J0339+099	&	54.8	&	10.0	&	10.3	&	2.5	&	0.2	&	ZwCl 0335+0956	&	54.647	&	9.965	&	Galaxy Cluster	&	0.0349	&	4U 0344+11	&		\\
21	&	MAXI J0343$-$537	&	55.9	&	$-$53.8	&	10.9	&	2.3	&	0.2	&	Abell 3158	&	55.665	&	$-$53.631	&	Galaxy Cluster	&	0.0597	&	4U 0339$-$54	&		\\
22	&	MAXI J0355+310	&	58.9	&	31.1	&	139.9	&	60.2	&	0.4	&	X Per                         	&	58.846	&	31.046	&	HMXB/NS              	&	 	&	4U 0352+30, 4U 0352+309	&		\\
23	&	MAXI J0413+105	&	63.4	&	10.5	&	18.7	&	4.7	&	0.2	&	Abell 478	&	63.336	&	10.476	&	Galaxy Cluster	&	0.0881	&	4U 0410+10	&		\\
24	&	MAXI J0423$-$569	&	65.9	&	$-$56.9	&	8.1	&	1.7	&	0.2	&	1RXS J042601.6$-$571202	&	66.503	&	$-$57.200	&	Sy1.5                	&	0.104	&		&	\\
25	&	MAXI J0431$-$613	&	68.0	&	$-$61.4	&	20.1	&	4.2	&	0.2	&	Abell 3266                    	&	67.800	&	$-$61.406	&	Galaxy Cluster       	&	0.0589	&	4U 0427$-$61	&		\\
26	&	MAXI J0433+053	&	68.4	&	5.3	&	11.7	&	2.8	&	0.2	&	3C 120                        	&	68.296	&	5.354	&	Sy1                  	&	0.03301	&	4U 0432+05	&		\\
27	&	MAXI J0433$-$131	&	68.5	&	$-$13.2	&	15.3	&	3.6	&	0.2	&	Abell 496	&	68.308	&	$-$13.259	&	Galaxy Cluster	&	0.0329	&	4U 0431$-$12	&		\\
28	&	MAXI J0457$-$696	&	74.5	&	$-$69.7	&	8.1	&	1.5	&	0.2	&	UNKNOWN	&	---	&	---	&	---	&	 --- 	&		&		\\
29	&	MAXI J0510+166	&	77.7	&	16.7	&	11.5	&	2.8	&	0.2	&	IRAS 05078+1626               	&	77.690	&	16.499	&	Sy1.5                	&	0.01788	&	4U 0517+17	&	\\
30	&	MAXI J0514$-$400	&	78.5	&	$-$40.1	&	37.5	&	11.7	&	0.3	&	NGC 1851                      	&	78.528	&	$-$40.044	&	LMXB/NS in globular c	&	 	&		&		\\
31	&	MAXI J0516$-$001	&	79.1	&	$-$0.2	&	9.4	&	2.2	&	0.2	&	Ark 120                       	&	79.048	&	$-$0.150	&	Sy1                  	&	0.0323	&	H 0523$-$00	&		\\
32	&	MAXI J0520$-$719	&	80.1	&	$-$71.9	&	100.9	&	27.0	&	0.3	&	LMC X-2                       	&	80.117	&	$-$71.965	&	LMXB                 	&	 	&	4U 0520$-$72	&		\\
33	&	MAXI J0529$-$327	&	82.3	&	$-$32.7	&	15.5	&	4.1	&	0.3	&	TV Col                        	&	82.356	&	$-$32.818	&	CV/DQ Her            	&	 	&	3A 0527$-$329	&		\\
34	&	MAXI J0533$-$662	&	83.3	&	$-$66.3	&	33.5	&	7.5	&	0.2	&	LMC X-4                       	&	83.207	&	$-$66.371	&	HMXB/NS              	&	 	&	4U 0532$-$66, 4U 0532$-$664	&		\\
35	&	MAXI J0535$-$054	&	83.8	&	$-$5.5	&	18.2	&	4.8	&	0.3	&	Trapezium Cluster             	&	83.819	&	$-$5.387	&	Star Cluster         	&	 	&	1ES 0532$-$05.4	&	\\
36	&	MAXI J0539$-$640	&	84.8	&	$-$64.1	&	89.0	&	25.4	&	0.3	&	LMC X-3                       	&	84.734	&	$-$64.082	&	HMXB/BHC	&	 	&	4U 0538$-$64	&		\\
37	&	MAXI J0539$-$696	&	85.0	&	$-$69.6	&	69.9	&	17.0	&	0.2	&	LMC X-1                       	&	84.911	&	$-$69.743	&	HMXB/BHC	&	 	&	4U 0540$-$69	&		\\
38	&	MAXI J0542+606	&	85.7	&	60.7	&	11.3	&	2.5	&	0.2	&	BY Cam                        	&	85.704	&	60.859	&	CV/AM Her            	&	 	&	4U 0541+60	&		\\
39	&	MAXI J0550$-$322	&	87.6	&	$-$32.2	&	8.3	&	2.1	&	0.2	&	PKS 0548$-$322                  	&	87.670	&	$-$32.272	&	BL Lac               	&	0.069	&	4U 0543$-$31	&		\\
40	&	MAXI J0552$-$075	&	88.1	&	$-$7.5	&	29.6	&	8.1	&	0.3	&	NGC 2110                      	&	88.047	&	$-$7.456	&	Sy2                  	&	0.00779	&	1H 0551$-$074	&		\\
41	&	MAXI J0555+464	&	88.8	&	46.5	&	15.1	&	3.7	&	0.2	&	MCG +08-11-011                	&	88.723	&	46.439	&	Sy1.5                	&	0.02048	&	4U 0558+46	&		\\
42	&	MAXI J0627$-$540	&	96.8	&	$-$54.0	&	14.4	&	3.2	&	0.2	&	ABELL 3391	&	96.564	&	$-$53.681	&	Galaxy Cluster	&	0.0531	&	1ES 0625$-$53.6	&	\\
''	&	''	&	''	&	''	&	''	&	''	&	''	&	ABELL 3395	&	96.879	&	$-$54.399	&	Galaxy Cluster	&	0.0498	&	1E 0626.2$-$5428	&	\\
43	&	MAXI J0710+590	&	107.7	&	59.1	&	9.0	&	2.1	&	0.2	&	QSO B0706+591	&	107.625	&	59.139	&	BL Lac               	&	0.125	&	1H 0658+595	&		\\
44	&	MAXI J0744$-$532	&	116.1	&	$-$53.3	&	7.5	&	1.6	&	0.2	&	V0436 Car	&	116.243	&	$-$52.952	&	CV/DQ Her	&	 	&	RX J0744.9$-$5257	&		\\
45	&	MAXI J0814$-$572	&	123.7	&	$-$57.3	&	9.4	&	1.9	&	0.2	&	CIZA J0812.5$-$5714	&	123.126	&	$-$57.235	&	Galaxy Cluster	&	0.062	&		&	\\
46	&	MAXI J0817$-$074	&	124.3	&	$-$7.5	&	9.6	&	2.5	&	0.3	&	Abell 644	&	124.356	&	$-$7.517	&	Galaxy Cluster	&	0.0704	&	2A 0815$-$075	&		\\
47	&	MAXI J0909$-$097	&	137.4	&	$-$9.7	&	21.3	&	5.8	&	0.3	&	Abell 754                     	&	137.209	&	$-$9.637	&	Galaxy Cluster       	&	0.0542	&	4U 0900$-$09	&		\\
48	&	MAXI J0918$-$121	&	139.6	&	$-$12.2	&	7.2	&	1.7	&	0.2	&	Abell 780	&	139.527	&	$-$12.093	&	Galaxy Cluster	&	0.0539	&	1H 0917$-$121	&		\\
49	&	MAXI J0924$-$317	&	141.0	&	$-$31.7	&	21.2	&	5.8	&	0.3	&	6dFGS gJ092353.7$-$314131     	&	140.974	&	$-$31.692	&	Sy1.9                	&	0.04225	&	SWIFT J0924.2$-$3141	&		\\
50	&	MAXI J0947$-$309	&	146.9	&	$-$31.0	&	24.1	&	6.6	&	0.3	&	MCG $-$05-23-016                	&	146.917	&	$-$30.949	&	Sy2                  	&	0.00849	&	4U 0945$-$30	&		\\
51	&	MAXI J0957+693	&	149.3	&	69.4	&	11.2	&	2.2	&	0.2	&	M81	&	148.888	&	69.065	&	Galaxy	&	 	&		&	\\
''	&	''	&	''	&	''	&	''	&	''	&	''	&	M82	&	148.970	&	69.679	&	Galaxy	&	 	&	4U 1004+70	&		\\
''	&	''	&	''	&	''	&	''	&	''	&	''	&	1RXSJ095755.0+690310	&	149.472	&	69.063	&	ULX	&	 	&	1E 0953.8+6918	&		\\
52	&	MAXI J1023+197	&	155.8	&	19.8	&	8.0	&	1.8	&	0.2	&	NGC 3227                      	&	155.877	&	19.865	&	Sy1.5                	&	0.00386	&	3A 1021+203	&		\\
53	&	MAXI J1036$-$276	&	159.2	&	$-$27.6	&	13.0	&	3.2	&	0.2	&	Abell 1060	&	159.214	&	$-$27.526	&	Galaxy Cluster	&	0.0126	&	4U 1033$-$26	&		\\
54	&	MAXI J1104$-$235	&	166.0	&	$-$23.5	&	7.2	&	1.6	&	0.2	&	1ES 1101$-$232	&	165.907	&	$-$23.492	&	BL Lac               	&	0.186	&	4U 1057$-$21	&		\\
55	&	MAXI J1104+382	&	166.1	&	38.2	&	89.3	&	33.5	&	0.3	&	Mrk 421                       	&	166.114	&	38.209	&	BL Lac               	&	0.03002	&	3A 1102+385	&	(E)	\\
56	&	MAXI J1105+725	&	166.3	&	72.5	&	9.2	&	1.8	&	0.2	&	NGC 3516                      	&	166.698	&	72.569	&	Sy1.5                	&	0.00884	&	1E 1103.4+7250	&		\\
57	&	MAXI J1135+676	&	174.0	&	67.6	&	7.5	&	1.5	&	0.2	&	RGB J1136+676	&	174.125	&	67.618	&	BL Lac               	&	0.13423	&	RX J1136.5+6737	&		\\
58	&	MAXI J1139$-$378	&	174.9	&	$-$37.8	&	12.3	&	3.2	&	0.3	&	NGC 3783                      	&	174.757	&	$-$37.739	&	Sy1                  	&	0.00973	&	4U 1136$-$37	&		\\
59	&	MAXI J1141+715	&	175.4	&	71.5	&	9.8	&	2.1	&	0.2	&	DO Dra                        	&	175.910	&	71.689	&	CV/DQ Her            	&	 	&	3A 1148+719	&		\\
60	&	MAXI J1144+198	&	176.2	&	19.8	&	11.8	&	2.7	&	0.2	&	Abell 1367	&	176.186	&	19.700	&	Galaxy Cluster	&	0.022	&	4U 1143+19	&		\\
61	&	MAXI J1144$-$184	&	176.2	&	$-$18.4	&	8.4	&	1.9	&	0.2	&	2MASX J11454045$-$1827149       	&	176.419	&	$-$18.454	&	Sy1                  	&	0.03295	&	1A 1143$-$18	&		\\
62	&	MAXI J1210+394	&	182.6	&	39.4	&	23.4	&	6.2	&	0.2	&	NGC 4151                      	&	182.636	&	39.406	&	Sy1.5                	&	0.00332	&	4U 1206+39	&		\\
63	&	MAXI J1218+038	&	184.7	&	3.8	&	8.0	&	2.0	&	0.2	&	ZwCl 1215.1+0400	&	184.419	&	3.662	&	Galaxy Cluster	&	0.075	&	1RXS J121742.6+033943 	&		\\
64	&	MAXI J1229+020	&	187.3	&	2.0	&	30.3	&	8.7	&	0.3	&	3C 273                        	&	187.278	&	2.052	&	Blazar	&	0.15834	&	4U 1226+02	&		\\
65	&	MAXI J1230+124	&	187.6	&	12.4	&	53.3	&	17.2	&	0.3	&	Virgo galaxy cluster	&	187.706	&	12.391	&	Galaxy Cluster	&	0.00436	&	H 1228+125 	&  (B)	\\
66	&	MAXI J1237$-$386	&	189.4	&	$-$38.6	&	7.4	&	1.8	&	0.3	&	V1025 Cen	&	189.569	&	$-$38.712	&	CV/DQ Her	&	 	&	RX J1238.2$-$3842	&		\\
67	&	MAXI J1240$-$052	&	190.1	&	$-$5.3	&	7.8	&	2.0	&	0.3	&	NGC 4593                      	&	189.914	&	$-$5.344	&	Sy1                  	&	0.009	&	4U 1240$-$05	&		\\
68	&	MAXI J1249$-$413	&	192.3	&	$-$41.3	&	28.2	&	8.4	&	0.3	&	Abell 3526	&	192.216	&	$-$41.306	&	Galaxy Cluster	&	0.0114	&	4U 1246$-$41	&		\\
69	&	MAXI J1252$-$292	&	193.2	&	$-$29.3	&	24.3	&	7.3	&	0.3	&	EX Hya                        	&	193.102	&	$-$29.249	&	CV/DQ Her            	&	 	&	4U 1249$-$28	&		\\
70	&	MAXI J1258$-$176	&	194.5	&	$-$17.6	&	8.7	&	2.1	&	0.2	&	Abell 1644	&	194.311	&	$-$17.354	&	Galaxy Cluster	&	0.0473	&	3A 1254$-$171	&	\\
71	&	MAXI J1259+279	&	194.9	&	28.0	&	70.1	&	22.5	&	0.3	&	Coma Cluster                  	&	194.953	&	27.981	&	Galaxy Cluster       	&	0.0231	&	4U 1257+28	&	(B)	\\
72	&	MAXI J1325$-$430	&	201.4	&	$-$43.0	&	76.8	&	28.7	&	0.4	&	Cen A                         	&	201.365	&	$-$43.019	&	Sy2                  	&	0.00183	&	4U 1322$-$42	&		\\
73	&	MAXI J1327$-$272	&	201.9	&	$-$27.2	&	7.7	&	1.9	&	0.2	&	Abell 1736                    	&	202.084	&	$-$27.299	&	Galaxy Cluster       	&	0.0458	&	1H 1326$-$269	&		\\
74	&	MAXI J1328$-$315	&	202.2	&	$-$31.6	&	18.7	&	5.1	&	0.3	&	Abell 3558	&	201.990	&	$-$31.503	&	Galaxy Cluster	&	0.048	&	1ES 1325$-$31.2	&	\\
75	&	MAXI J1335$-$342	&	203.9	&	$-$34.3	&	13.7	&	3.4	&	0.2	&	MCG $-$06-30-015                	&	203.974	&	$-$34.296	&	Sy1.2                	&	0.00775	&	H 1332$-$336	&		\\
76	&	MAXI J1336$-$237	&	204.0	&	$-$23.8	&	7.1	&	1.7	&	0.2	&	ESO 509-IG66	&	203.668	&	$-$23.446	&	Sy2                  	&	0.03431	&	IRAS 13319$-$2311	&  (F)	\\
77	&	MAXI J1338+045	&	204.6	&	4.6	&	11.8	&	2.9	&	0.2	&	NGC 5252                      	&	204.566	&	4.543	&	Sy1.9                	&	0.02297	&	SWIFT J1338.2+0433	&		\\
78	&	MAXI J1347$-$328	&	206.9	&	$-$32.9	&	24.8	&	7.1	&	0.3	&	Abell 3571                    	&	206.871	&	$-$32.866	&	Galaxy Cluster       	&	0.0391	&	1ES 1344$-$32.6	&		\\
79	&	MAXI J1348+267	&	207.2	&	26.7	&	16.3	&	3.9	&	0.2	&	Abell 1795	&	207.252	&	26.585	&	Galaxy Cluster	&	0.062476	&	4U 1348+25	&		\\
80	&	MAXI J1349$-$302	&	207.3	&	$-$30.3	&	25.9	&	7.4	&	0.3	&	IC 4329A                      	&	207.330	&	$-$30.309	&	Sy1.2                	&	0.01605	&	3A 1346$-$301	&		\\
81	&	MAXI J1358$-$480	&	209.7	&	$-$48.1	&	7.5	&	1.9	&	0.2	&	CIZA J1358.9$-$4750	&	209.700	&	$-$47.630	&	Galaxy Cluster	&	0.074	&	1RXS J135842.4$-$475029	&	\\
82	&	MAXI J1407$-$509	&	212.0	&	$-$50.9	&	8.5	&	1.9	&	0.2	&	CIZA J1407.8$-$5100	&	211.969	&	$-$51.009	&	Galaxy Cluster	&	0.0966	&	XSS J14080$-$5042	&	\\
83	&	MAXI J1413$-$031	&	213.3	&	$-$3.2	&	20.8	&	5.7	&	0.3	&	NGC 5506                      	&	213.312	&	$-$3.208	&	Sy1.9                	&	0.00618	&	4U 1410$-$03	&		\\
84	&	MAXI J1418+251	&	214.6	&	25.2	&	12.7	&	3.1	&	0.2	&	NGC 5548                      	&	214.498	&	25.137	&	Sy1.5                	&	0.01717	&	4U 1414+25	&		\\
85	&	MAXI J1429+426	&	217.4	&	42.6	&	8.9	&	2.1	&	0.2	&	1ES 1426+428                  	&	217.136	&	42.672	&	BL Lac               	&	0.129	&	2MASX J14283260+4240210, 4U 1444+43	&		\\
86	&	MAXI J1511+058	&	227.9	&	5.9	&	23.1	&	6.1	&	0.3	&	Abell 2029	&	227.733	&	5.745	&	Galaxy Cluster	&	0.07728	&	3A 1509+058	&	\\
87	&	MAXI J1516+065	&	229.2	&	6.6	&	7.9	&	2.1	&	0.3	&	Abell 2052	&	229.226	&	6.530	&	Galaxy Cluster	&	0.035491	&	1ES 1514+07.2	&	\\
88	&	MAXI J1522+279	&	230.7	&	28.0	&	8.4	&	2.0	&	0.2	&	Abell 2065	&	230.678	&	27.723	&	Galaxy Cluster	&	0.0726	&	4U 1521+28	&		\\
89	&	MAXI J1522+084	&	230.7	&	8.4	&	8.4	&	2.0	&	0.2	&	Abell 2063	&	230.772	&	8.602	&	Galaxy Cluster	&	0.03616	&	1ES 1520+08.7	&	\\
90	&	MAXI J1558+272	&	239.7	&	27.3	&	21.2	&	5.2	&	0.2	&	Abell 2142                    	&	239.567	&	27.225	&	Galaxy Cluster       	&	0.0909	&	4U 1556+27	&		\\
91	&	MAXI J1602+160	&	240.6	&	16.0	&	10.9	&	2.6	&	0.2	&	Abell 2147	&	240.572	&	15.895	&	Galaxy Cluster	&	0.035	&	4U 1601+15	&	\\
92	&	MAXI J1603$-$759	&	240.8	&	$-$75.9	&	8.5	&	1.5	&	0.2	&	CIZA J1601.7$-$7544	&	240.445	&	$-$75.746	&	Galaxy Cluster	&	0.153	&	XSS J16019$-$7548	&	\\
93	&	MAXI J1615$-$058	&	243.9	&	$-$5.8	&	8.8	&	2.3	&	0.3	&	Abell 2163	&	243.946	&	$-$6.146	&	Galaxy Cluster	&	0.1698	&	1H 1613$-$060	&		\\
94	&	MAXI J1617$-$283	&	244.5	&	$-$28.4	&	9.0	&	2.2	&	0.2	&	1RXS J161933.6$-$280736      	&	244.887	&	$-$28.127	&	Symb/NS              	&	 	&	SWIFT J1619.4$-$2808	&	(G)	\\
95	&	MAXI J1619$-$156	&	245.0	&	$-$15.7	&	2976.9	&	19216.8	&	6.2	&	Sco X-1                       	&	244.980	&	$-$15.640	&	LMXB/NS              	&	 	&	4U 1617$-$15	&		\\
96	&	MAXI J1626$-$335	&	246.7	&	$-$33.5	&	9.9	&	2.5	&	0.2	&	CIZA J1626.3$-$3329	&	246.586	&	$-$33.489	&	Galaxy Cluster	&	0.1098	&	XSS J16265$-$3303	&	\\
97	&	MAXI J1627$-$243	&	246.8	&	$-$24.4	&	11.1	&	2.8	&	0.2	&	3EG J1627$-$2419	&	247.000	&	$-$24.300	&	BL Lac	&	 --- 	&		&		\\
98	&	MAXI J1628+396	&	247.2	&	39.6	&	16.9	&	4.4	&	0.2	&	Abell 2199	&	247.154	&	39.524	&	Galaxy Cluster	&	0.030151	&	4U 1627+39	&		\\
99	&	MAXI J1632$-$674	&	248.0	&	$-$67.4	&	83.5	&	21.9	&	0.3	&	KZ TrA	&	248.070	&	$-$67.462	&	LMXB                 	&	 	&	4U 1626$-$67, 4U 1626$-$673	&		\\
100	&	MAXI J1633+056	&	248.3	&	5.6	&	8.1	&	2.0	&	0.2	&	Abell 2204	&	248.194	&	5.571	&	Galaxy Cluster	&	0.152158	&	2E 1630.3+0540	&		\\
101	&	MAXI J1633$-$750	&	248.3	&	$-$75.0	&	7.8	&	1.3	&	0.2	&	ABELL 3628	&	247.755	&	$-$75.176	&	Galaxy Cluster	&	0.1492	&	XSS J16345$-$7506	&	\\
''	&	''	&	''	&	''	&	''	&	''	&	''	&	1RXS J163153.5$-$750631	&	247.973	&	$-$75.109	&	X-ray source	&	 --- 	&		&		\\
102	&	MAXI J1638$-$643	&	249.6	&	$-$64.3	&	41.9	&	9.4	&	0.2	&	TrA Cluster                   	&	249.567	&	$-$64.347	&	Galaxy Cluster       	&	0.0508	&	XSS J16384$-$6424	&		\\
103	&	MAXI J1653+398	&	253.4	&	39.8	&	13.2	&	3.2	&	0.2	&	Mrk 501                       	&	253.468	&	39.760	&	BL Lac               	&	0.03366	&	4U 1651+39	&		\\
104	&	MAXI J1657+353	&	254.5	&	35.4	&	45.4	&	29.3	&	0.4	&	Her X-1                       	&	254.458	&	35.342	&	LMXB/NS              	&	 	&	4U 1656+35	&		\\
105	&	MAXI J1701+341	&	255.3	&	34.2	&	9.2	&	2.5	&	0.3	&	Abell 2244	&	255.679	&	34.062	&	Galaxy Cluster	&	0.0968	&	3A 1702+340	&		\\
106	&	MAXI J1703+785	&	256.0	&	78.6	&	18.0	&	4.0	&	0.2	&	Abell 2256                    	&	255.931	&	78.718	&	Galaxy Cluster       	&	0.0581	&	4U 1707+78	&	\\
107	&	MAXI J1706+241	&	256.7	&	24.1	&	10.0	&	2.3	&	0.2	&	V934 Her	&	256.644	&	23.972	&	LMXB/NS              	&	 	&	4U 1700+24	&		\\
108	&	MAXI J1716$-$629	&	259.2	&	$-$62.9	&	8.8	&	1.6	&	0.2	&	NGC 6300                      	&	259.248	&	$-$62.821	&	Sy2                  	&	0.0037	&	SWIFT J1717.1$-$6249	&		\\
109	&	MAXI J1741+185	&	265.5	&	18.5	&	7.3	&	1.7	&	0.2	&	4C+18.51	&	265.530	&	18.456	&	Sy1	&	0.186	&	RX J1742.1+1827	&		\\
110	&	MAXI J1753$-$014	&	268.4	&	$-$1.4	&	130.1	&	64.2	&	0.5	&	SWIFT J1753.5$-$0127            	&	268.368	&	$-$1.452	&	LMXB/ BHC            	&	 	&		&		\\
111	&	MAXI J1807+061	&	271.9	&	6.1	&	7.7	&	1.9	&	0.2	&	V0426 Oph	&	271.966	&	5.862	&	CV/Dwarf Nova	&	 	&	1RXS J180751.8+055144	&	\\
112	&	MAXI J1815+498	&	274.0	&	49.8	&	31.3	&	8.7	&	0.3	&	AM Her                        	&	274.055	&	49.868	&	CV/AM Her            	&	 	&	4U 1813+50, 4U 1814+50	&		\\
113	&	MAXI J1825$-$371	&	276.4	&	$-$37.1	&	76.9	&	32.4	&	0.4	&	V0691 CrA	&	276.445	&	$-$37.105	&	LMXB                 	&	 	&	4U 1822$-$371, 4U 1822$-$37	&		\\
114	&	MAXI J1826+305	&	276.5	&	30.6	&	7.9	&	1.8	&	0.2	&	CIZA J1825.3+3026	&	276.353	&	30.442	&	Galaxy Cluster	&	0.065	&		&		\\
115	&	MAXI J1835+328	&	278.8	&	32.8	&	11.5	&	2.7	&	0.2	&	3C 382                        	&	278.764	&	32.696	&	Sy1                  	&	0.05787	&	H 1832+325	&		\\
116	&	MAXI J1835$-$328	&	278.9	&	$-$32.9	&	23.0	&	7.3	&	0.3	&	XB 1832$-$330 (NGC 6652)	&	278.934	&	$-$32.991	&	LMXB/NS              	&	 	&	NGC 6652	&		\\
117	&	MAXI J1837$-$653	&	279.5	&	$-$65.3	&	12.0	&	2.3	&	0.2	&	ESO 103-G35                   	&	279.585	&	$-$65.428	&	Sy2                  	&	0.01329	&	H 1834$-$653	&		\\
118	&	MAXI J1839+798	&	279.9	&	79.9	&	13.6	&	2.8	&	0.2	&	3C 390.3                      	&	280.538	&	79.771	&	Sy1                  	&	0.0561	&	4U 1847+78	&		\\
119	&	MAXI J1851$-$783	&	282.9	&	$-$78.3	&	8.5	&	1.5	&	0.2	&	2MASX J18470283$-$7831494       	&	281.762	&	$-$78.530	&	Sy1                  	&	0.07412	&	4U 1916$-$79	&	\\
120	&	MAXI J1855$-$311	&	283.8	&	$-$31.2	&	21.4	&	6.5	&	0.3	&	V1223 Sgr                     	&	283.759	&	$-$31.163	&	CV/DQ Her            	&	 	&	4U 1849$-$31	&		\\
121	&	MAXI J1900$-$249	&	285.0	&	$-$24.9	&	68.8	&	27.2	&	0.4	&	HETE J1900.1$-$2455             	&	285.036	&	$-$24.921	&	LMXB/msPSR           	&	 	&		&		\\
122	&	MAXI J1920$-$586	&	290.1	&	$-$58.6	&	11.8	&	2.5	&	0.2	&	ESO 141-G55                	&	290.309	&	$-$58.670	&	Sy1                  	&	0.036	&	4U 1924$-$59	&		\\
123	&	MAXI J1921+439	&	290.3	&	44.0	&	35.5	&	9.7	&	0.3	&	Abell 2319                    	&	290.189	&	43.962	&	Galaxy Cluster       	&	0.0557	&	4U 1919+44	&		\\
124	&	MAXI J1924+501	&	291.0	&	50.2	&	10.1	&	2.4	&	0.2	&	CH Cyg                        	&	291.138	&	50.241	&	Symb/WD              	&	 	&	1RXS J192433.0+501415	&		\\
125	&	MAXI J1941$-$104	&	295.4	&	$-$10.4	&	14.1	&	3.8	&	0.3	&	NGC 6814	&	295.669	&	$-$10.323	&	Sy1	&	0.005214	&	2E 1939.9$-$1026	&	\\
''	&	''	&	''	&	''	&	''	&	''	&	''	&	V1432 Aql	&	295.048	&	$-$10.424	&	CV/AM Her            	&	 	&	RX J1940.1$-$1025	&		\\
126	&	MAXI J1959+650	&	299.9	&	65.1	&	20.3	&	4.9	&	0.2	&	1ES 1959+650             	&	299.999	&	65.148	&	BL Lac               	&	0.047	&	QSO B1959+650	&		\\
127	&	MAXI J2009$-$611	&	302.3	&	$-$61.1	&	8.0	&	1.6	&	0.2	&	NGC 6860                      	&	302.195	&	$-$61.100	&	Sy1                  	&	0.01488	&	RX J200847$-$61055	&		\\
128	&	MAXI J2011$-$568	&	302.9	&	$-$56.9	&	22.0	&	5.2	&	0.2	&	Abell 3667                    	&	303.125	&	$-$56.816	&	Galaxy Cluster       	&	0.0556	&	1ES 2008$-$57.0	&		\\
129	&	MAXI J2041+750	&	310.4	&	75.0	&	9.3	&	1.8	&	0.2	&	4C +74.26                     	&	310.655	&	75.134	&	Sy1                  	&	0.104	&	XSS J20404+7521	&		\\
130	&	MAXI J2044$-$107	&	311.0	&	$-$10.7	&	12.3	&	3.3	&	0.3	&	Mrk 509                       	&	311.041	&	$-$10.723	&	Sy1.2                	&	0.0344	&	3A 2041$-$107	&		\\
131	&	MAXI J2129+121	&	322.5	&	12.2	&	42.7	&	12.2	&	0.3	&	4U 2129+12                    	&	322.493	&	12.167	&	LMXB in globular clus	&	 	&	4U 2127+119	&		\\
132	&	MAXI J2135$-$626	&	323.8	&	$-$62.7	&	7.4	&	1.5	&	0.2	&	1RXS J213623.1$-$622400         	&	324.096	&	$-$62.400	&	Sy1                  	&	0.0588	&	6dFGS gJ213623.1$-$622401	&		\\
133	&	MAXI J2144+383	&	326.2	&	38.3	&	550.6	&	702.3	&	1.2	&	Cyg X-2                       	&	326.172	&	38.322	&	LMXB/NS              	&	 	&	4U 2142+38	&		\\
134	&	MAXI J2201$-$600	&	330.3	&	$-$60.1	&	8.3	&	1.7	&	0.2	&	Abell 3827	&	330.483	&	$-$59.949	&	Galaxy Cluster	&	0.0984	&	1RXS J220157.8$-$595648	&		\\
135	&	MAXI J2202$-$319	&	330.6	&	$-$31.9	&	7.6	&	2.0	&	0.3	&	NGC 7172                      	&	330.508	&	$-$31.870	&	Sy2                  	&	0.00868	&	SWIFT J2201.8$-$3152	&		\\
136	&	MAXI J2214+125	&	333.6	&	12.6	&	7.2	&	1.7	&	0.2	&	RU Peg                        	&	333.511	&	12.703	&	CV/Dwarf Nova	&	 	&	2E 2211.5+1227	&		\\
137	&	MAXI J2218$-$084	&	334.6	&	$-$8.4	&	12.5	&	3.3	&	0.3	&	FO Aqr   	&	334.481	&	$-$8.351	&	CV/DQ Her            	&	 	&	2E 2215.3$-$0835	&		\\
138	&	MAXI J2235$-$259	&	339.0	&	$-$25.9	&	9.3	&	2.5	&	0.2	&	NGC 7314                      	&	338.943	&	$-$26.050	&	Sy1.9                	&	0.00476	&	3A 2233$-$259	&		\\
139	&	MAXI J2253+163	&	343.5	&	16.3	&	17.0	&	4.4	&	0.3	&	3C 454.3                      	&	343.491	&	16.148	&	Blazar	&	0.859	&	1E 2251.5+1553	&		\\
140	&	MAXI J2253$-$177	&	343.5	&	$-$17.7	&	11.8	&	3.0	&	0.2	&	MR 2251$-$178                   	&	343.524	&	$-$17.582	&	Sy1                  	&	0.06398	&	2A 2251$-$179 	&		\\
141	&	MAXI J2255$-$031	&	343.8	&	$-$3.1	&	12.7	&	3.4	&	0.3	&	AO Psc                        	&	343.825	&	$-$3.178	&	CV/DQ Her            	&	 	&	3A 2253$-$033	&		\\
142	&	MAXI J2305$-$085	&	346.3	&	$-$8.6	&	14.8	&	3.9	&	0.3	&	Mrk 926                       	&	346.181	&	$-$8.686	&	Sy1.5                	&	0.04686	&	4U 2305$-$07	&		\\
143	&	MAXI J2316$-$425	&	349.1	&	$-$42.6	&	7.1	&	2.0	&	0.3	&	Abell S1101	&	348.494	&	$-$42.727	&	Galaxy Cluster	&	0.058	&	1E 2311.2$-$4259	&	\\

\enddata
\tablenotetext{a}{The flux is in units of 10$^{-11}$ ergs cm$^{-2}$ s$^{-1}$ in the 4--10 keV band, converted from Crab units; 1 Crab = 1.21$\times$10$^{-8}$ ergs cm$^{-2}$ s$^{-1}$.}
\tablenotetext{b}{Sy: Seyfert galaxy, LMXB: low-mass X-ray binary, HMXB: high-mass X-ray binary, Symb: symbiotic star, NS: neutron star, BHC: black hole candidate, CV: cataclysmic variable, ULX: ultra-luminous X-ray source, msPSR: millisecond pulsar.}
\tablenotetext{c}{The redshifts are taken from the online database SIMBAD and NED. A blank indicates the source is Galactic, and --- indicates the source has no redshift information.}
\tablenotetext{d}{(A) Identified by individual inspection since in the
image fitting the position is unable to be determined due to significant
source confusion with Perseus Cluster.\ (B) The data is extended
compared with the PSF model.\ (C) Matsumura et al.(2011)\ (D)Suzuki et
al.(2010)\ (E) Isobe et al.(2010)\ (F) This source is likely to be 
contaminated by 1RXS J133344.4$-$231734 and TV Hya. 
ESO 509-IG66 has double nuclei with similar fluxes in the
2--10 keV band (Guainazzi et al. 2005). The redshift is adopted from
Sekiguchi \& Wolstencroft (1992).\ (G) This source is possibly contaminated 
by V893 Sco and other nearby sources.}

\end{deluxetable}
\renewcommand{\arraystretch}{1.0}

\end{landscape}

%% file: ref_catalog.tex
\begin{table*}
  \begin{center}
    \caption{Reference catalogs used for cross correlation.}
    \begin{tabular}{lcl}
      \hline\hline
      Catalog & Number of & Reference\\
              & matched sources & \\
      \hline
      Swift/BAT 58-month Catalog   &  93 & \cite{Baumgartner2010}$^*$ \\
      54-month Palermo BAT-Survey Catalog (2PBC) & 100 & \cite{2010A&A...524A..64C} \\
      1st Fermi/LAT AGN Catalog (1LAC)     &  12 & \cite{2010ApJ...715..429A} \\
      RASS Bright Souce Catalog (1RXS) & 110 & \cite{1999A&A...349..389V} \\
      INTEGRAL General Reference Catalog (ver. 31) & 108 & INTEGRAL Science Data Centre website$^{\dagger}$ \\
      NORAS Galaxy Cluster Survey Catalog &  14 &\cite{2000ApJS..129..435B} \\
      REFLEX Galaxy Cluster Survey Catalog &  17 & \cite{2004A&A...425..367B} \\
      1st XMM-Newton Slew Survey Catalog (XMMSL1) &  65 & \cite{2008A&A...480..611S} \\
      \hline
      \multicolumn{3}{@{}l@{}}{\hbox to 0pt{\parbox{160mm}{\footnotesize
            \par\noindent
            \footnotemark[$*$] The data is retrieved from NASA website: http://heasarc.nasa.gov/docs/swift/results/bs58mon
            \par\noindent
            \footnotemark[$\dagger$] http://www.isdc.unige.ch/integral/science/catalogue
            %
          }\hss}}
      \label{tab:ref_catalog}
    \end{tabular}
  \end{center}
\end{table*}

%% file: type-table.tex
\begin{table}
  \begin{center}
    \caption{Categories of cataloged sources.}
    \begin{tabular}{ccc}
      \hline\hline
      Category & & Number of sources\\
      \hline
      unidentified     & &  1 \\
      galaxies         & &  1 \\
      galaxy clusters  & & 48 \\
      Seyfert galaxies  & & 39 \\
      blazars          & & 12 \\
      CVs/Stars        & & 20 \\
      X-ray binaries   & & 18 \\
      confused & &  4 \\
      \hline
      \label{tab:category}
    \end{tabular}
  \end{center}
\end{table}